\documentclass[aps,prd,twocolumn,floatfix,nofootinbib,showpacs,superscriptaddress,tightenlines]{revtex4-1}
\usepackage{amsmath}
\usepackage{lipsum}
\usepackage{amssymb}
\usepackage{amsthm}
\usepackage{bbold}
\usepackage{dcolumn}
\usepackage{epsfig}
\usepackage{graphics}
\usepackage{graphicx}
\usepackage{longtable}
\usepackage{color}
\usepackage{epstopdf}
\usepackage{xspace}
\usepackage{cancel}
\usepackage{nicefrac}
\usepackage[colorlinks=true, pdfstartview=FitV, linkcolor=purple, citecolor= purple,urlcolor=blue]{hyperref}
\usepackage{lipsum}

\definecolor{darkgreen}{rgb}{0,0.5,0}
\definecolor{purple}{rgb}{0.5,0,0.5}
\definecolor{nblue}{rgb}{0.0,0.0,0.50}
\definecolor{scarlet}{rgb}{1.0,0.2,0}

\newcommand{\M}{\scriptsize{\hbox{M}}}
\em
\begin{document}

\title{Pseudo-scalar mesons: light front wave functions, GPDs and PDFs}

\author{L. Albino}
\affiliation{Instituto de F\'{i}sica y Matem\'aticas, Universidad
Michoacana de San Nicol\'as de Hidalgo, Morelia, Michoac\'an
58040, M\'{e}xico.}
\affiliation{Instituto de F\'isica Te\'orica, Universidade Estadual Paulista, Rua Dr. Bento Teobaldo Ferraz, 271-Bloco II, 01140-070, S\~{a}o Paulo, SP, Brazil.}
\author{I. M. Higuera-Angulo}
\affiliation{Instituto de F\'{i}sica y Matem\'aticas, Universidad
Michoacana de San Nicol\'as de Hidalgo, Morelia, Michoac\'an
58040, M\'{e}xico.}
\author{K. Raya}
\affiliation{Department of Integrated Sciences and Center for Advanced Studies in Physics, Mathematics and Computation, University of Huelva, E-21071 Huelva, Spain.}
\affiliation{Departamento de F\'sica Te\'orica y del Cosmos, Universidad de Granada, E-18071, Granada, Spain. }
\author{A. Bashir}
\affiliation{Instituto de F\'{i}sica y Matem\'aticas, Universidad
Michoacana de San Nicol\'as de Hidalgo, Morelia, Michoac\'an
58040, M\'{e}xico.}

\date{\today}

\begin{abstract}

   We develop a unified algebraic model which satisfactorily describes the internal structure of pion and kaon as well as heavy quarkonia ($\eta_c$ and $\eta_b$). For each of these mesons, we compute their generalized parton distributions (GPDs), built through the overlap representation of their light-front wave function, tightly constrained by the modern and precise knowledge of their quark distribution amplitudes. From this three-dimensional knowledge of mesons, we deduce parton distribution functions (PDFs) as well as electromagnetic form factors and construct the impact parameter space GPDs. The PDFs for mesons formed with light quarks are then evolved from the hadronic scale of around 0.3 GeV to 5.2 GeV, probed in experiments. We make explicit comparisons with experimental results available and with earlier theoretical predictions.

\end{abstract}


\maketitle
\date{\today}

\section{Introduction}
\label{SECTION Introduction}

The generalized parton distributions (GPDs) have
emerged as a comprehensive tool to describe hadron
structure probed in hard scattering 
processes~\cite{Radyushkin:1996nd,Ji:1996ek,Diehl:2000xz,Burkardt:2002hr,Diehl:2003ny,Belitsky:2005qn,Guidal:2013rya,Constantinou:2020hdm}. GPDs
connect hadron electromagnetic form factors (FFs)~\cite{Guidal:2004nd}, measured in elastic 
processes, to longitudinal parton
distributions (PDFs) which are probed in deep inelastic scattering~\cite{Ellis:1996mzs}. 
They provide 
a kaleidoscopic view of the three-dimensional
spatial structure of hadrons, written as a function of the longitudinal momentum
fraction $x$, the momentum transfer $t$ and the  skewness variable $\xi$ (longitudinal momentum transfer); furthermore, Fourier transform
of the GPDs yields the transverse spatial distribution of
partons correlated with $x$~\cite{Burkardt:2000za}, the so called impact parameter space GPDs (IPS-GPDs).

Most physical observables related to mesons can be
calculated through a combined knowledge of their Bethe-Salpeter amplitude (BSA)
and the quark propagator~\cite{Roberts:1994dr,Eichmann:2016yit}. While in principle it can accurately be achieved through a cumbersome 
computation of the quark propagator Schwinger-Dyson equation (SDE) and the Bethe-Salpeter equation (BSE) in close connection with
full QCD~\cite{Qin:2020rad}, calculation of a plethora of  experimentally interesting quantities such as FFs~\cite{Chang:2013nia,Raya:2015gva,Raya:2016yuj,Ding:2018xwy,Gao:2017mmp,Eichmann:2019bqf,Miramontes:2021exi,Raya:2022ued}, distribution amplitudes (PDAs), PDFs~\cite{Chang:2013pq,Ding:2019lwe,Ding:2019qlr,Cui:2020dlm,Cui:2020tdf,Cui:2021mom,Cui:2022bxn}, and, specially, GPDs~\cite{Raya:2022eqa,Raya:2021zrz,Zhang:2021mtn,Chavez:2021koz,Chavez:2021llq}, remains a highly non-trivial pursuit. However, our understanding of the intricate interplay between the quark propagator and the meson BSA  permits us to construct their simple {\em Anst$\ddot{a}$ze} sufficiently efficacious to make reliable predictions and amicable enough to offer algebraic manipulations. 
In this article, we carry out this algebraic model (AM) construction for
pseudo-scalar mesons in terms of a form-invariant spectral density. The most attractive feature of this AM is that the spectral density is explicitly written in terms of the leading-twist PDA whose existing reliable information allows us to {\em circumvent the need to construct any ad hoc Ansatz for the spectral density.}

 We begin with an evidence-based {\em Ansatz} for the quark propagator and the BSA in terms of a spectral density function which is form-invariant for all the ground-state pseudo-scalar mesons. 
 The Bethe-Salpeter wave function (BSWF) can then be readily constructed whose subsequent projection on to the light front yields the highly sought-after light-front wave function (LFWF). Its integration over the transverse momentum squared ($k_{\bot}^2$) gives us access to the valence-quark PDA. We exploit our current detailed and accurate knowledge of the PDAs of pseudo-scalar mesons~\cite{Cui:2020tdf,Ding:2015rkn} to determine the parameters of our model. We use the overlap representation of the LFWF~\cite{Diehl:2000xz} to compute the GPDs of pion, kaon, $\eta_c$ and $\eta_b$. From this three dimensional knowledge of these mesons, different limits/projections lead us to deduce
 the PDFs, FFs and the IPS-GPDs which are then compared to available experimental extractions of these observables.
 
 This article has been organized as follows. In Sec. II, we present our generalized AM for the quark propagator and the BSA of the pseudo-scalar mesons under consideration in terms of a spectral density function. It allows us to derive the leading-twist
 LFWF in Sec. III by merely appealing to the definition of its Mellin moments. The resulting LFWF permits establishing a closed and simple algebraic connection with the PDA, so that the need to specify a spectral density is completely avoided if the PDA is known.
 Sec. IV details the extraction of GPDs through
 the overlap representation of the LFWFs as suggested in~\cite{Diehl:2000xz}, in the so called DGLAP kinematic region, and a series of distributions derived therefrom: PDFs, FFs and IPS-GPDs. In Sec. V, we particularize our AM to produce a collection of distributions, using inputs from previous SDE predictions, and compare (when possible) with available theoretical calculations and experimental data. Finally, in Sec. VI, we present a summary of our work and the scope of our model.

\section{Algebraic Model}
\label{SECTION Algebraic model}
The internal dynamics of a meson can be described, in a fully quantum field theoretic formalism, via its BSWF. In terms of the associated BSA ($\Gamma_{\M}$), and the quark (antiquark) propagators ($S_{q,(\bar{h})}$), the BSWF reads
\begin{eqnarray}
{\chi}_{\M} \left(k_{-}, P \right) = S_q(k) \Gamma_{\M}\left(k_{-}, P\right) S_{\bar{h}}\left(k-P\right) \,,
\label{BS1}
\end{eqnarray}
where $k_{-}=k-P/2$ and $P^2=-m_{\M}^2$ is the (negative) mass squared of the meson $\M$. The labels $q$ and $\bar{h}$ which denote the valence quark and antiquark flavors are in general different but might also be the same. Although the propagators and BSA might be obtained from solutions of the corresponding SDEs and BSEs, useful and relevant insight can be extracted from sensibly constructed simpler models. 

Plain expressions for the quark (antiquark) propagator and BSAs that capture QCD's key non-perturbative traits are given by
\begin{eqnarray}
S_{q(\bar{h})}(k) &=& \left[-i\gamma \cdot k + M_{q(\bar{h})}\right] \Delta\left(k^2,M_{q(\bar{h})}^2\right) 
\,,
\label{Anzats1}\\
n_{\M} \Gamma_{\M}(k,P) &=& i\gamma_5 \int_{-1}^{1} dw \, \rho_{\M}(w)\left[ \hat{\Delta} \left(k_w^2 , \Lambda_w^2 \right) \right]^\nu \hspace{-.1cm},
\label{Anzats2}
\end{eqnarray}
where $\Delta(s,t)=(s+t)^{-1},\;\hat{\Delta}(s,t) = t \Delta(s,t),  k_w = k + (w/2)P$. Herein, $M_{q(\bar{h})}$ is a mass scale akin to a constituent mass for a given quark flavor and $n_{\M}$ is a normalization constant that will be determined later. The function $\rho_{\M}(w)$ can be regarded as a spectral density, whose particular form determines the pointwise behavior of the BSA, therefore having a crucial impact on the meson observables. The parameter $\nu > -1$ controls the asymptotic behavior of the BSA; this is discussed in detail below. Finally, $\Lambda_w^2 \equiv \Lambda^2(w)$ is defined as follows:
\begin{eqnarray}
\Lambda^2(w) & = & M_{q}^2-\frac{1}{4}\left(1-w^2\right)m_{\M}^2 \nonumber \\ 
&& +\frac{1}{2} \left( 1 -w  \right)\left(M_{\bar{h}}^2-M_q^2\right) \,.
\label{Lambda}
\end{eqnarray}
Notice that, unlike kindred models~\cite{ Chavez:2021koz,Chavez:2021llq,Chouika:2017dhe,Chouika:2017rzs, Mezrag:2016hnp, Mezrag:2014jka,Raya:2022eqa,Raya:2021zrz,Zhang:2021mtn,Xu:2018eii} which have been employed successfully to compute an array of GPD-related distributions, we have promoted $\Lambda \to \Lambda_w$ to incorporate a $w$-dependence.  Keeping in mind the efficacy of earlier models, we point out some key differences which yield a simplification of relevant integrals and  closed algebraic expressions relating different distributions:

 \begin{itemize}
     \item We retain the constant term from the original models, setting it to $M_q$. 
     \item There is a term linear in $w$ which is the only term not symmetric under $w \leftrightarrow -w$. This asymmetry allows us to study mesons with different flavored quarks and is hence accompanied with the multiplicative factor of $( {M_{\overline{h}}}^2 - M_q^2 )$. When quark-antiquark are of the same flavor, this term ceases to contribute by construction. 
     
     \item Then we have a quadratic term in $w^2$ which has the coefficient proportional to $m_M^2$. We choose the coefficients of each power of $w$ to ensure that the condition:
\end{itemize}
\vspace{-0.8cm}
\begin{eqnarray}
|M_{\bar{h}}-M_q| \leq m_{\M} \leq M_{\bar{h}}+M_q 
\label{mass inequalities}
\end{eqnarray}
guarantees the positivity of  $\Lambda^2(w)$.

When quark and antiquark have the same flavor, the left part of the inequality is trivially satisfied. We consider isospin symmetry, i.e., $M_u = M_d$. In any other case, like that of a kaon or heavy-light mesons, the ratio $M_{\bar{h}}/M_{q}$ must be set with care. Realistic solutions of the quark SDE provide useful benchmarks~\cite{AtifSultan:2018end}. 
Note that $m_{\M} < M_{\bar{h}}+M_q$ is satisfied for Nambu-Goldstone bosons. One can find sensible values for the constituent masses to uphold this inequality for the ground state pseudo-scalar mesons.

Combining Eqs.~\eqref{BS1}-\eqref{Anzats2}, the BSWF acquires the following Nakanishi integral representation (NIR):
\begin{eqnarray}
n_{\M} \chi_{\M}(k_{-},P) \hspace{-.1cm} = \hspace{-.1cm}\mathcal{M}_{q,\bar{h}}(k,P) \hspace{-.1cm} \int_{-1}^1 \hspace{-.2cm} dw \, \tilde{\rho}^{\,\nu}_{\M}(w)\mathcal{D}^{\,\nu}_{q,\bar{h}}(k,P) \,,
\label{BSA}
\end{eqnarray}
\begin{figure}[!t]
    \centering
    \includegraphics[scale=.24]{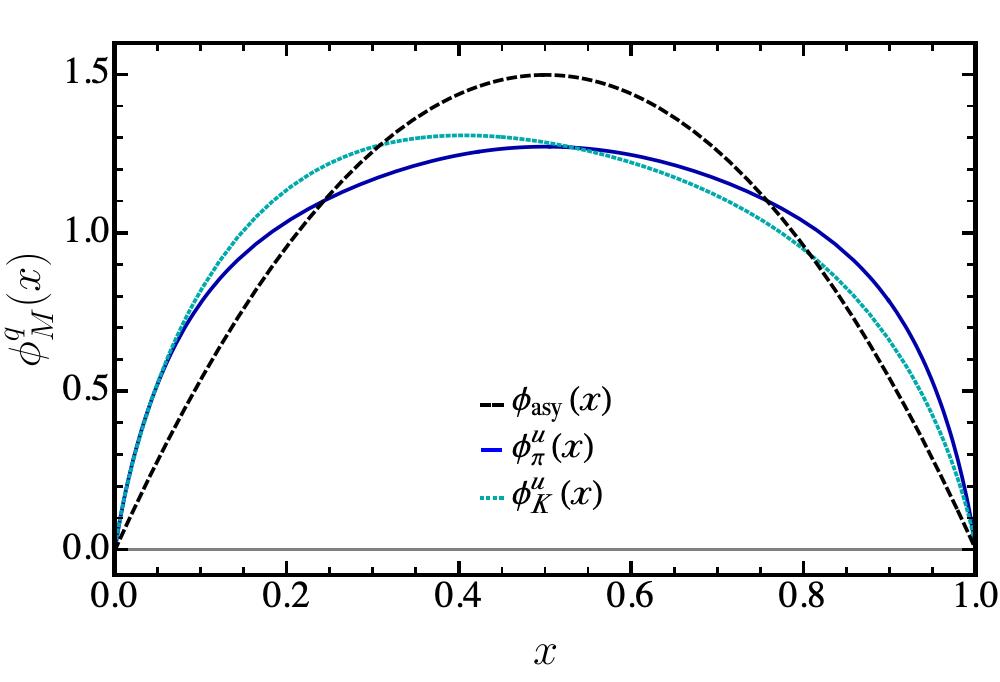}
    \includegraphics[scale=.24]{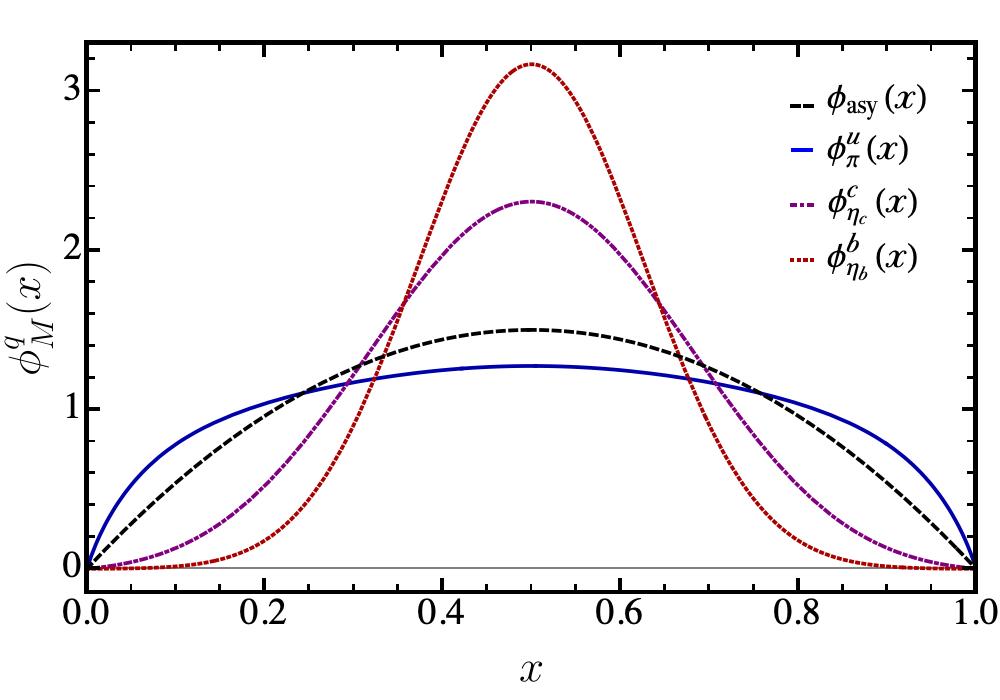}
    \caption{Upper panel- pion and kaon PDAs at $\zeta_H$. Lower panel- the corresponding ones for $\eta_c$ and $\eta_b$. The distributions were obtained within the SDE formalism in Refs.~\cite{Cui:2020tdf,Ding:2015rkn}, and parameterized according to Eqs.~\eqref{eq:PDAsSDE}. For comparison, the asymptotic distribution, $\phi_{asy}(x)=6x(1-x)$, is also shown.}
    \label{fig:PDA1}
\end{figure}
\begin{figure*}[!ht]
    \centering
    \includegraphics[width=0.45\textwidth]{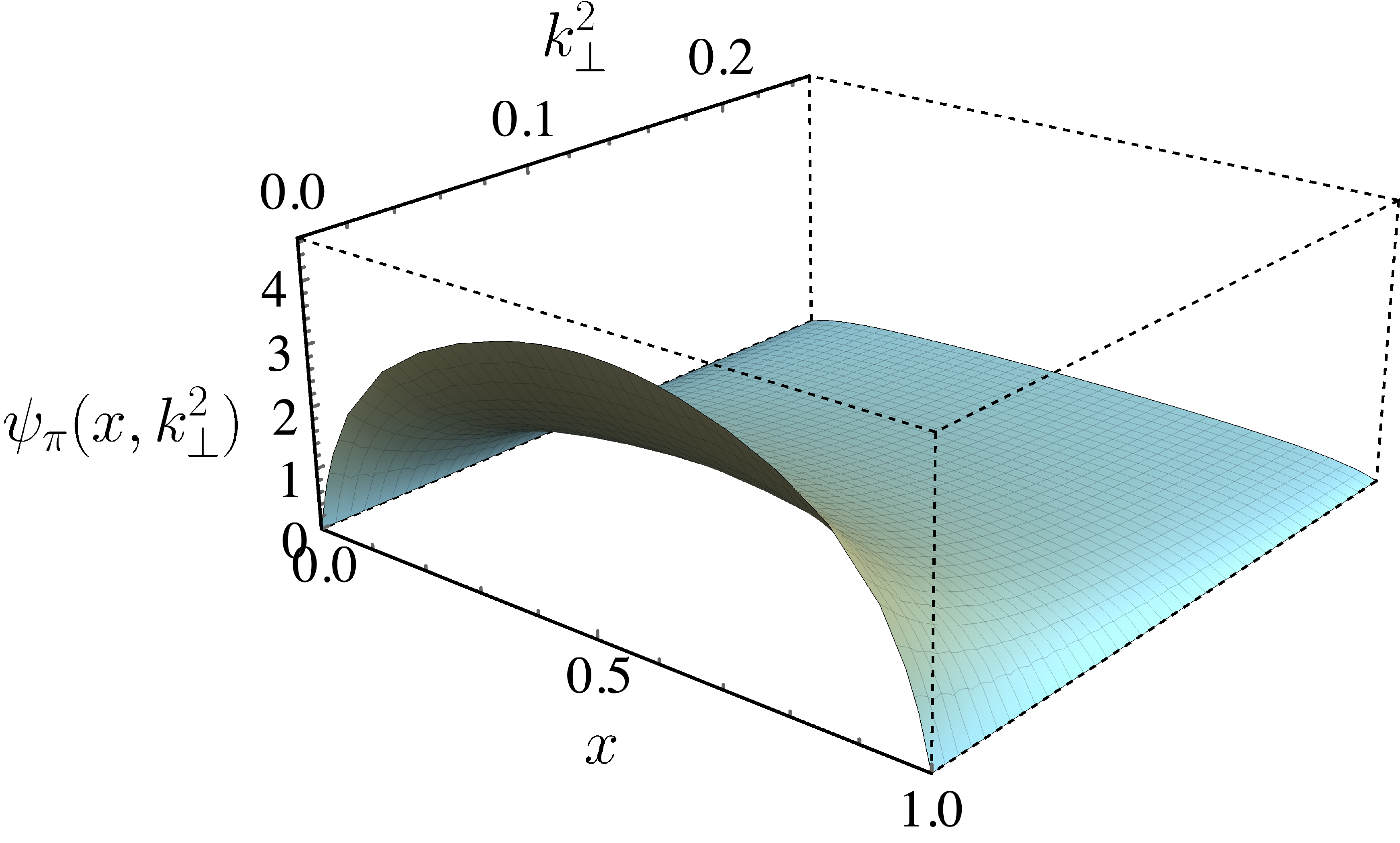}
    \includegraphics[width=0.45\textwidth]{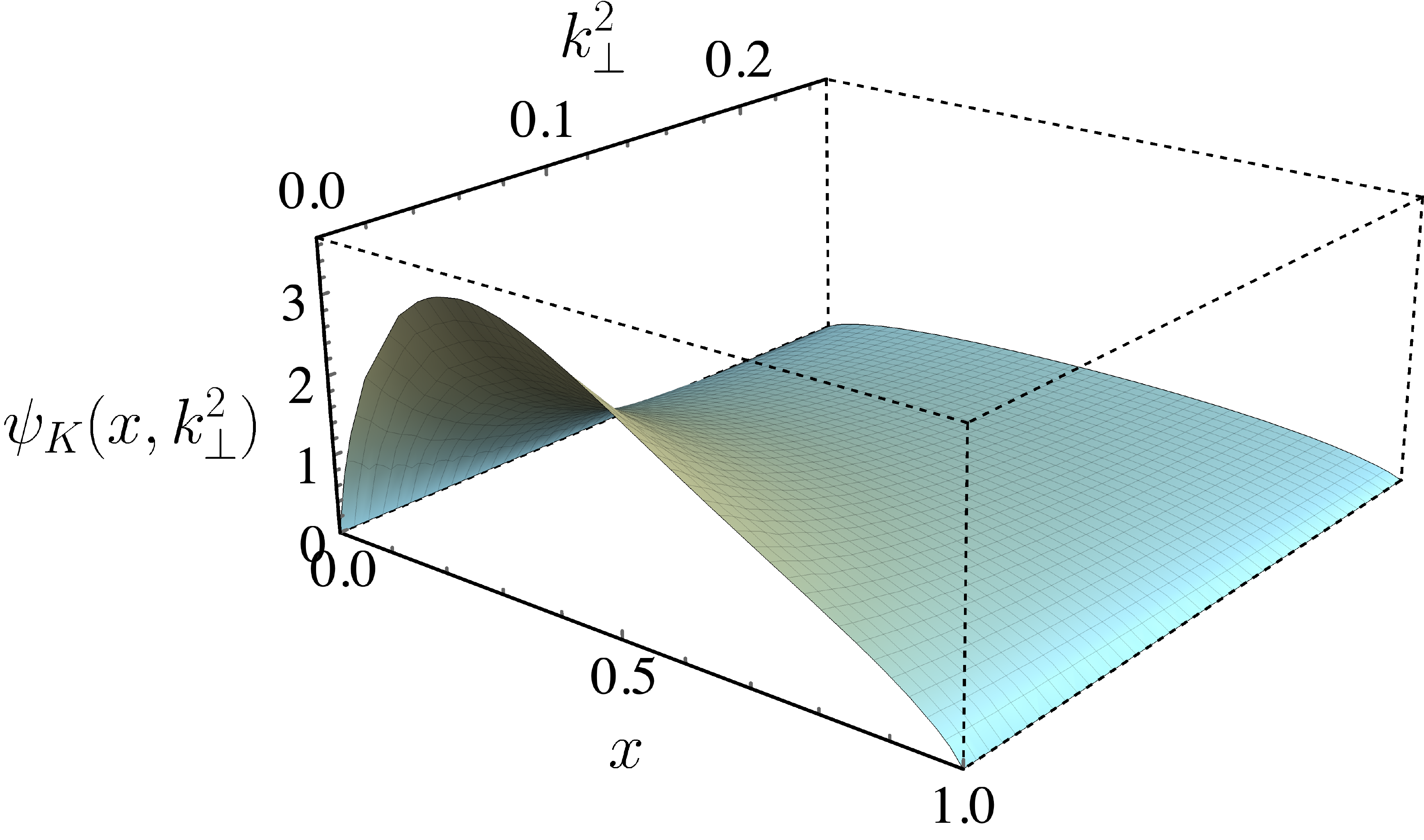} \\
    \includegraphics[width=0.45\textwidth]{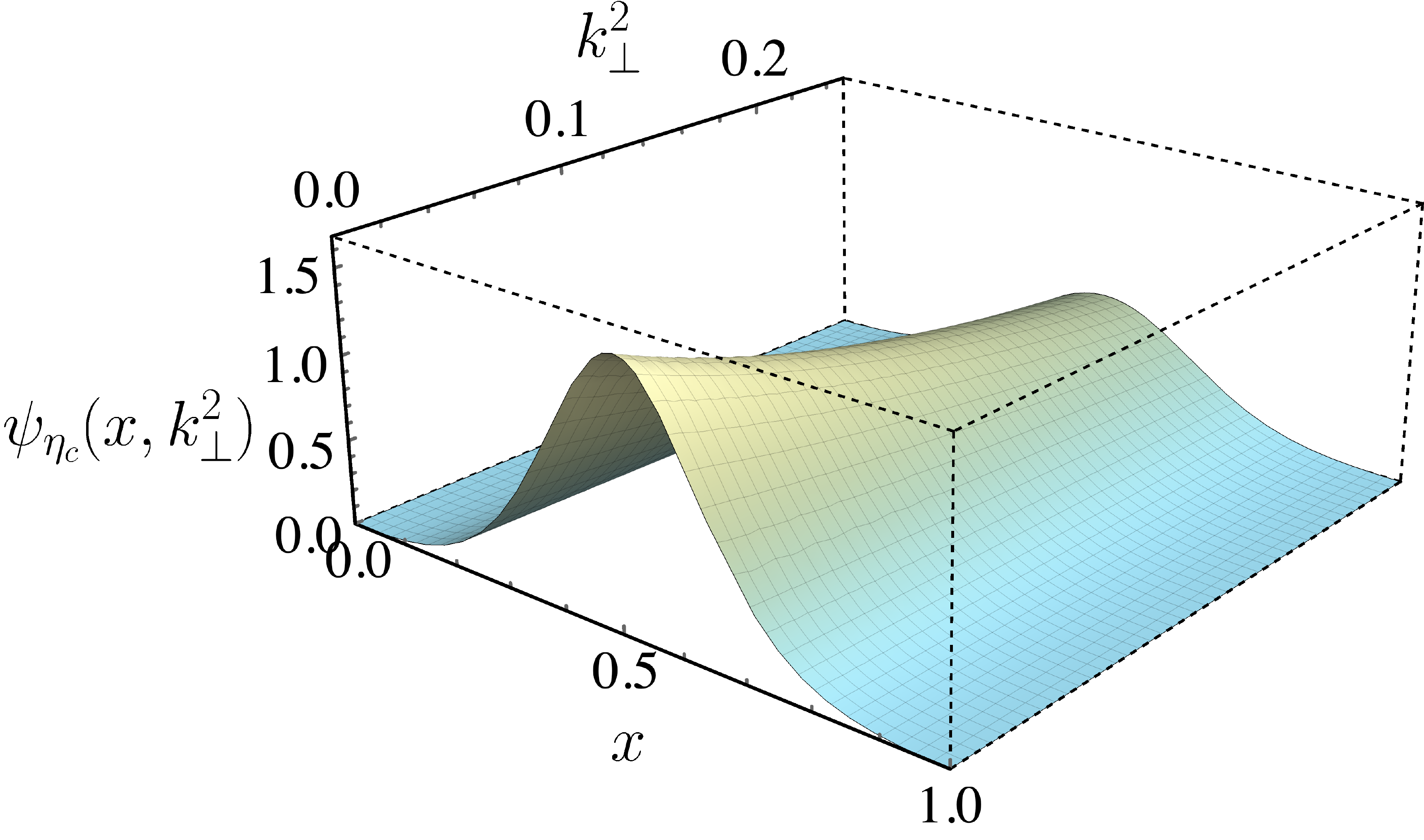}
    \includegraphics[width=0.45\textwidth]{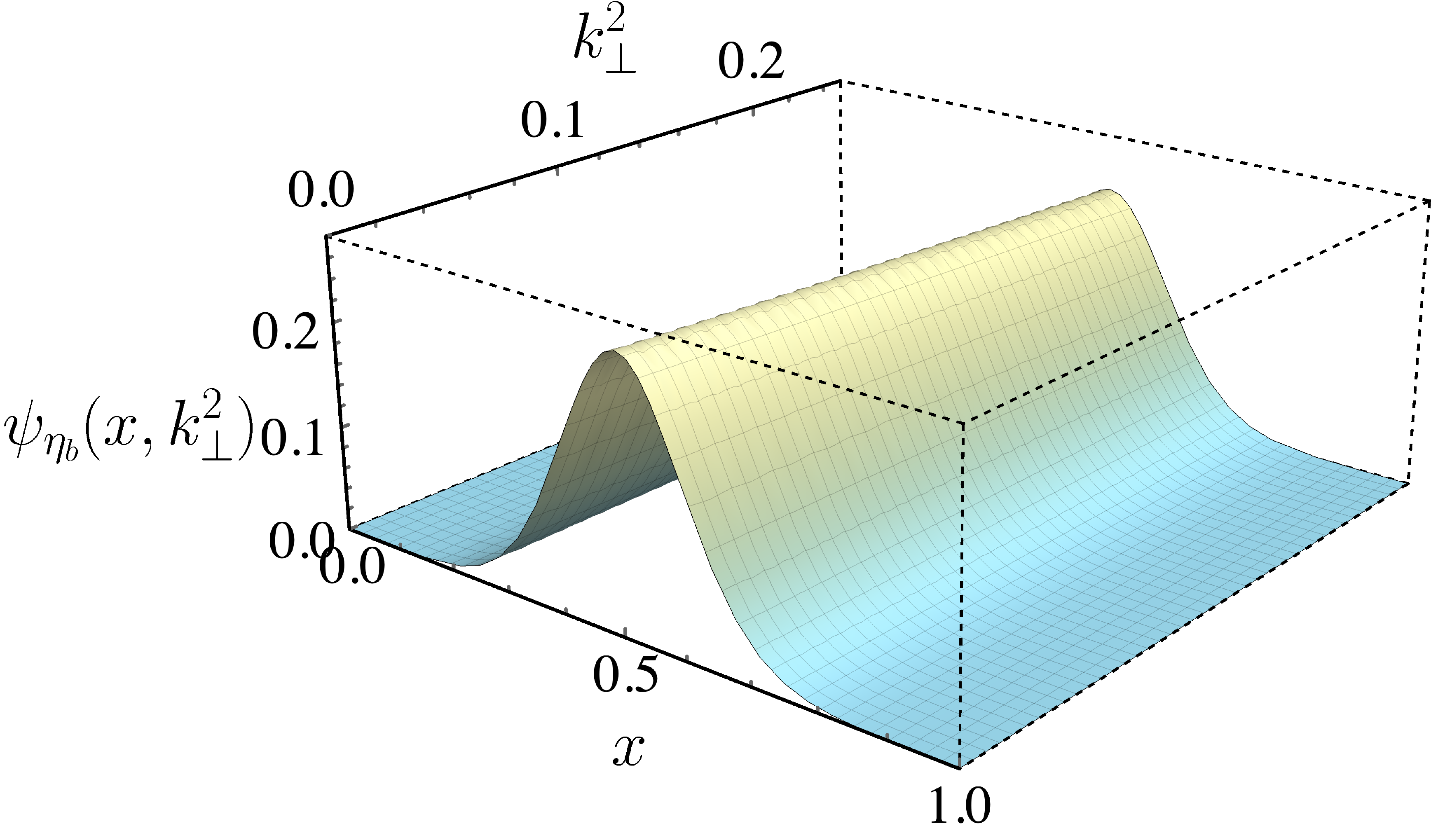}
    \caption{LFWFs of the pion, kaon, $\eta_c$ and $\eta_b$ obtained from Eq.~(\ref{eq:LFWFPDArel}) and the inputs described in Section V. Herein we have depicted $\psi_{\M}(x,k_\perp^2) \to \psi_{\M}(x,k_\perp^2)/(16\pi^2 f_{\M})$. Mass units in GeV.}
    \label{PionLFWF}
\end{figure*}
\hspace{-2mm}where the profile function, $\tilde{\rho}^{\,\nu}_{\M}(w)$, has been defined in terms of the spectral density as
\begin{eqnarray}
\tilde{\rho}^{\,\nu}_{\M}(w) \equiv \rho_{\M}(w)\Lambda_w^{2\nu} \,.
\label{Effective Rho Definition}
\end{eqnarray}
The function $\mathcal{M}_{q,\bar{h}}(k=p+P,P)$ has the following tensor structure that also characterizes $\chi_{M}(k,P)$:
\begin{eqnarray}
\hspace{-.5cm} \mathcal{M}_{q,\bar{h}}(k,P) \hspace{-.1cm} & \equiv & \hspace{-.1cm} -\gamma_5 \big[ M_q\gamma \cdot P + \gamma \cdot k (M_{\bar{h}}-M_q) \nonumber \\ 
&& \hspace{.5cm} + \sigma_{\mu\nu}k_{\mu}P_{\nu}  - i \, (k\cdot p + M_q M_{\bar{h}}) \big] \,. \label{M(k,P)}
\end{eqnarray}
Due to the trace over Dirac indices, \textit{cf.} eq. (\ref{LFWF}), the last two terms containing an even number of $\gamma$-matrices in the above eq. (\ref{M(k,P)}) do  not contribute to the leading-twist light-front wave function (LFWF) and, consequently, the PDA. The function $\mathcal{D}_{q,\bar{h}}^{\,\nu}(k,P)$ is a product of quadratic denominators,
\begin{eqnarray}\nonumber
\hspace{-.5cm} \mathcal{D}_{q,\bar{h}}^{\,\nu}(k,P) \hspace{-.1cm} & \equiv & \hspace{-.1cm} \Delta \left( k^2,M_q^2 \right) \nonumber \\ 
&&\hspace{.5cm} \times \Delta \left( k_{w-1}^2,\Lambda_w^2 \right)^{\nu}  \Delta \left( p^2,M_{\bar{h}}^2 \right) \,.\label{eq:dens}
\end{eqnarray}
Feynman parametrization enables us to combine the denominators in Eq.~\eqref{eq:dens} into a single one. Then a suitable change of variables and a subsequent rearrangement in the order of integration yields the expression:
\begin{eqnarray}
\label{eq:chits}
n_{\M}\chi_{\M} (k_{-},P)&=& \mathcal{M}_{q,\bar{h}}(k,P)  \int_0^1 d\alpha  \mathcal{F}_{\M}(\alpha,\sigma^{\nu+2})\;,\\   \label{eq:F0}
\mathcal{F}_{\M}(\alpha,\sigma^{\nu+2})&=&\nu(\nu+1)
 \Big[ \int_{-1}^{1-2\alpha} dw 
\int_{ \frac{2\alpha}{w-1} + 1}^1 d\beta \label{BSE2.1} \\
&& \hspace{-0.7 cm} + \int_{1-2\alpha}^1 dw \int_{ \frac{ 2\alpha+(w-1)}{w+1 } }^1 d\beta \Big] \frac{(1-\beta)^{\nu-1} \tilde{\rho}_{\M}^{\,\nu}(w) }{ \sigma^{\nu+2} } \,,\nonumber
\end{eqnarray}
where $\sigma =  (k-\alpha P)^2 + \Lambda^2_{1-2\alpha}$, and $\alpha,\;\beta$ are Feynman parameters. Since only $(1-\beta)^{\nu-1}$ depends on $\beta$, integration over $d\beta$ can be performed directly, thus yielding
\begin{eqnarray}
    \label{eq:FM1}
    \mathcal{F}_{\M}(\alpha,\sigma^{\nu+2})&=&2^\nu(\nu+1)\Big[ \int_{-1}^{1-2\alpha} dw \left(\frac{\alpha}{1-w}\right)^\nu\\
    &+&\int_{1-2\alpha}^1 dw \left(\frac{1-\alpha}{1+w}\right)^\nu\Big] \frac{\tilde{\rho}_{\M}^{\,\nu}(w)}{\sigma^{\nu+2}}\;.\nonumber
\end{eqnarray}
As we explain in the Appendix, this extra algebraic integration allows us to completely derive $\tilde{\rho}^{\,\nu}_{\M}(w)$ in terms of the PDA. In the next section we shall explicitly see that, when employing this model for the BSWF, many quantities and relations of interest can be obtained in a purely analytical manner.

\section{Light front wave functions and parton distribution amplitudes}
\label{SECTION Light front wave function}

For a quark $q$ within a pseudo-scalar meson $\M$, the leading twist (2 particle) light-front wave function, $\psi^q_{\M}$, can be obtained via the light-front projection of the meson's BSA as:
\begin{eqnarray}
\hspace{-0.3cm} \psi_{\M}^q \left( x,k_{\perp}^2 \right) = \text{tr} \int_{ dk_{\parallel} }\delta_n^x(k_{\M}) \gamma_5 \gamma \cdot  n \, \chi_{\M}(k_{-},P) \,,
\label{LFWF}
\end{eqnarray}
where $\delta_n^x(k_{\M})=\delta( n  \cdot  k - x \, n \cdot  P )$; $ n $ is a light-like four-vector, such that $ n^2 = 0 $ and $ n\cdot P = -m_{\M} $;  a mentioned before, $x$ corresponds to the light-front momentum fraction carried by the quark. The trace is taken over color and Dirac indices. The notation $\int_{dk_\parallel} \equiv \int \frac{d^2 k_\parallel}{\pi}$ has been employed and the 4-momentum integral is defined as usual:
\begin{eqnarray}
\int \frac{d^4k}{(2\pi)^4} = \left[ \frac{1}{16 \pi^3} \int d^2 k_\perp \right] \left[ \frac{1}{\pi} \int d^2 k_\parallel \right]\;.\;
\end{eqnarray}
The moments of the distribution are:
\begin{eqnarray}\label{eq:MellinMoments}
\langle x^m \rangle_{ \psi_{\M}^q } &=& \int_0^1 dx \, x^m \, \psi_{\M}^q \left( x,k_{\perp}^2 \right) \\
&=& \text{tr} \frac{1}{n\cdot P} \int_{dk_{\parallel}} \left[\frac{n\cdot k}{n\cdot P} \right]^m  \gamma_5 \gamma \cdot n \chi_{\M}(k_{-},P)\;.\nonumber
\end{eqnarray}
From Eqs.~\eqref{eq:chits}-\eqref{eq:MellinMoments}, one arrives at
\begin{eqnarray}
\langle x^m \rangle_{ \psi_{\M} ^q} &=& \int_0^1 d\alpha \alpha^m \left[\frac{12}{n_{\M}} \frac{\mathcal{Y}_{\M}(\alpha,\sigma_\perp^{\nu+1})}{\nu+1}\right]\;,
\label{LFWF1}\\
\mathcal{Y}_{\M}(\alpha,\sigma_\perp^{\nu+1})&=&\mathcal{F}_{\M}(\alpha,\sigma_\perp^{\nu+1})(\alpha M_{\bar{h}}+(1-\alpha)M_q)\,,\nonumber
\end{eqnarray}
where $\sigma_\perp = k_\perp^2+ \Lambda_{1-2\alpha}^2$. Uniqueness of the Mellin moments, Eqs.~\eqref{eq:MellinMoments}-\eqref{LFWF1}, implies the connection between the Feynman parameter $\alpha$ and the momentum fraction $x$; therefore one can identify the LFWF as
\begin{equation}
    \label{eq:LFWFgood}
    \psi_{\M}^q(x,k_\perp^2)=\left[\frac{12}{n_{\M}} \frac{\mathcal{Y}_{\M}(x,\sigma_\perp^{\nu+1})}{\nu+1}\right]\;.
\end{equation}
Notice that the above expression resembles the one derived, for instance, in~\cite{Xu:2018eii,Raya:2022eqa,Raya:2021zrz}. However, the crucial difference is the $w$-dependent definition of $\Lambda_w$, Eq.~\eqref{Lambda}. As mentioned before, its particular form enables additional simplicity and allows amicable algebraic manipulation as will be evident shortly.


Integrating out the $k_\perp$ dependence of $\psi_{\M}^q(x,k_\perp)$ yields the PDA,
\begin{eqnarray}
f_{\M}\phi_{\M}^q(x) = \frac{ 1}{16\pi^3 } \int d^2 k_{\perp} \psi_{\M}^q \left( x,k_{\perp}^2 \right) \,,
\label{eq:PDAdefinition}
\end{eqnarray}
where $f_{\M}$ is the leptonic decay constant of the meson. From Eqs.~\eqref{eq:FM1} and \eqref{eq:LFWFgood}, it is seen that the only term in the above equation that depends on $k_\perp$ is $1/\sigma_\perp^{\nu+1}$, then

\begin{eqnarray}\nonumber
    \frac{1}{16\pi^3}\int  d^2k_{\perp}\frac{1}{\sigma_\perp^{\nu+1}} &=& \frac{1}{8\pi^2}\int dk_\perp \frac{k_\perp}{(k_\perp^2+\Lambda_{1-2\alpha}^2)^{\nu+1}} \\
    &=& \frac{1}{16\pi^2}\frac{1}{\nu \Lambda_{1-2\alpha}^{2\nu}}\;.
\label{eq:cos0}
\end{eqnarray}
Combining Eqs.~\eqref{eq:LFWFgood}-\eqref{eq:cos0} we arrive at the following algebraic relation between $\psi_{\M}(x,k_\perp^2)$ and $\phi_{\M}(x)$:
\begin{equation}
    \label{eq:LFWFPDArel}
    \psi_{\M}^q(x,k_\perp^2) = 16\pi^2 f_{\M}\frac{\nu \Lambda_{1-2x}^{2\nu}}{(k_\perp^2+\Lambda_{1-2x}^2)^{\nu+1}}\phi_{\M}^q(x)\;.
\end{equation}
 The compact result above is a merit of the AM we have put forward. Throughout this manuscript, we shall employ dimensionless and unit normalized PDAs, $\int_0^1 dx \phi_{\M}^q(x)=1$. The resulting PDA and LFWF are expressed in a quasiparticle basis at an intrinsic scale, intuitively identified with some hadronic scale, $\zeta_H$, for which the valence degrees of freedom fully express the properties of the hadron under study. Most results herein are quoted at $\zeta_H$ (unless specified otherwise). However, for the sake of simplicity, the label $\zeta_H$ shall be omitted. It is worth reminding that the quark and antiquark PDA are connected via momentum conservation,
 \begin{eqnarray}
 \label{eq:PDAquarkantiquark}
 \phi_{\M}^q(x; \zeta_H)=\phi_{\M}^{\bar{h}}(1-x;\zeta_H) \;,
 \end{eqnarray}
a constricted and firm connection that prevails even after evolution~\cite{Lepage:1979zb,Efremov:1979qk,Lepage:1980fj}.
 
Some practical corollaries of the AM and Eq.~\eqref{eq:LFWFPDArel}:
\begin{itemize}
    \item Given a particular form of $\phi_{\M}^q(x)$, the $\psi_{\M}^q(x,k_\perp^2)$ can be obtained quite straightforwardly.
    \item As long as we have reliable access to $\phi_{\M}^q(x)$, there is no actual need to construct the profile function $\tilde{\rho}^{\,\nu}(w)$ (although it can be properly identified, as we explain in the Appendix). 
   \item It also works the other way around. A sensible choice of $\tilde{\rho}_{\M}^{\nu}(w)$ and model parameters yields algebraic expressions for both $\phi_{\M}^q(x)$ and $\psi_{\M}^q(x,k_{\perp}^2)$.
    \item In fact, the present AM can be reduced to the toy model employed in~Refs.~\cite{Chouika:2017rzs, Mezrag:2016hnp, Mezrag:2014jka} with appropriate substitutions. It also faithfully reproduces the results obtained from the more sophisticated {\em Ansatz} in Ref.~\cite{Raya:2022eqa,Raya:2021zrz,Zhang:2021mtn}.
    \item The degree of factorizability of the LFWF is clearly exposed through Eqs.~\eqref{Lambda} and \eqref{eq:LFWFPDArel}.
\end{itemize}
Regarding the last point let us consider the chiral limit ($m_{\M}=0$, $M_{q}=M_{\bar{h}}$), then $\Lambda_{1-2x}^2 = M_q^2$ and 
\begin{eqnarray}
    \label{eq:LFWFfact}
    \psi_{\M}^q(x,k_\perp^2) &=&  \left[16\pi^2f_{\M} \frac{\nu M_q^{2\nu}}{(k_\perp^2+M_q^2)^{\nu+1}}\right] \phi_{\M}^q(x)\;.
\end{eqnarray}
The bracketed term no longer depends on $x$; hence, the $x$ and $k_\perp$ dependence of $\psi_{\M}(x,k_\perp^2)$ has been completely factorized. Conversely, as captured by Eq.~\eqref{eq:LFWFPDArel}, a non-zero meson mass and quark/antiquark flavor asymmetry, namely $m_{\M}^2 \neq 0$ and $(M_{\bar{h}}^2-M_q^2) \neq 0$, yield a LFWF which correlates $x$ and $k_\perp^2$. So one should expect an increasingly dominant role of $x$ and $k_\perp^2$ correlations in heavy-quarkonia and heavy-light systems. Notably, a soft $Q^2$-dependence might also be introduced in the definition of the PDA~\cite{Rinaldi:2022dyh, Brodsky:2011yv}, Eq.~\eqref{eq:PDAdefinition}, producing the following compact expression:
\begin{eqnarray}
    \phi(x;Q^2) &=& \left(1- \frac{\Lambda^{2\nu}_{1-2x}}{ \left[Q^2+\Lambda^2_{1-2x}\right]^{\nu}} \right) \phi(x) \\ &\overset{\underset{\mathrm {\nu=1}}{}}{\rightarrow}& \left( \frac{Q^2}{Q^2+\Lambda^2_{1-2x}} \right) \phi(x)\,.\nonumber
\end{eqnarray}
Clearly,  $\phi(x;Q^2 \to \infty) = \phi(x)$, which is the limit we take for the sake of the discussion. In the next section, we shall exploit the virtues of Eq.~\eqref{eq:LFWFPDArel} to compute the pseudo-scalar meson GPDs in the overlap representation. 

\begin{figure*}[!ht]
    \centering
    \includegraphics[scale=.24]{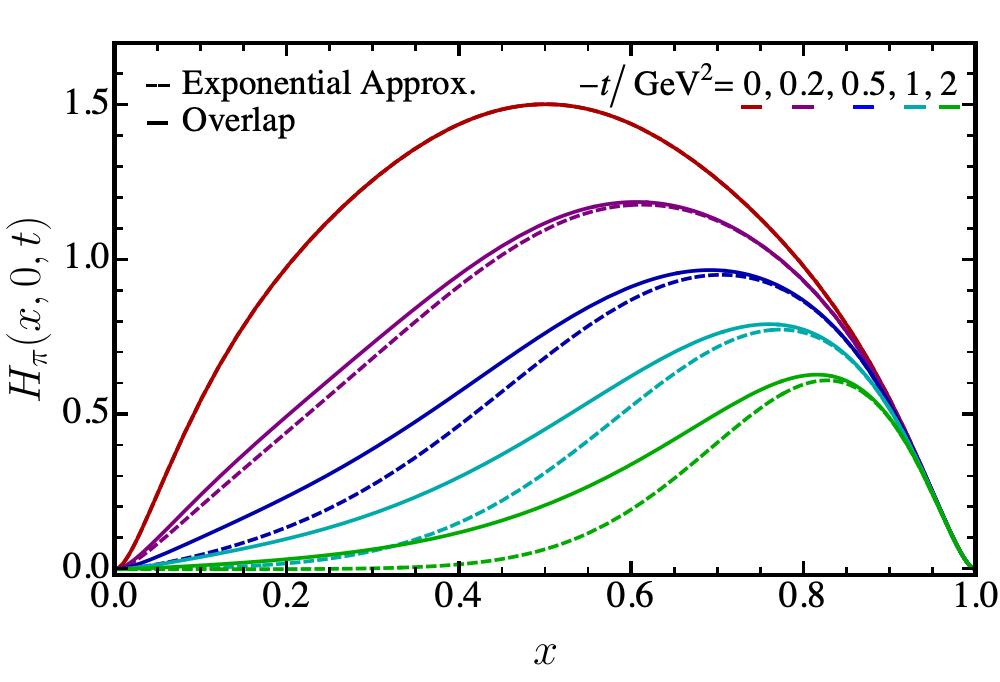}
    \includegraphics[scale=.24]{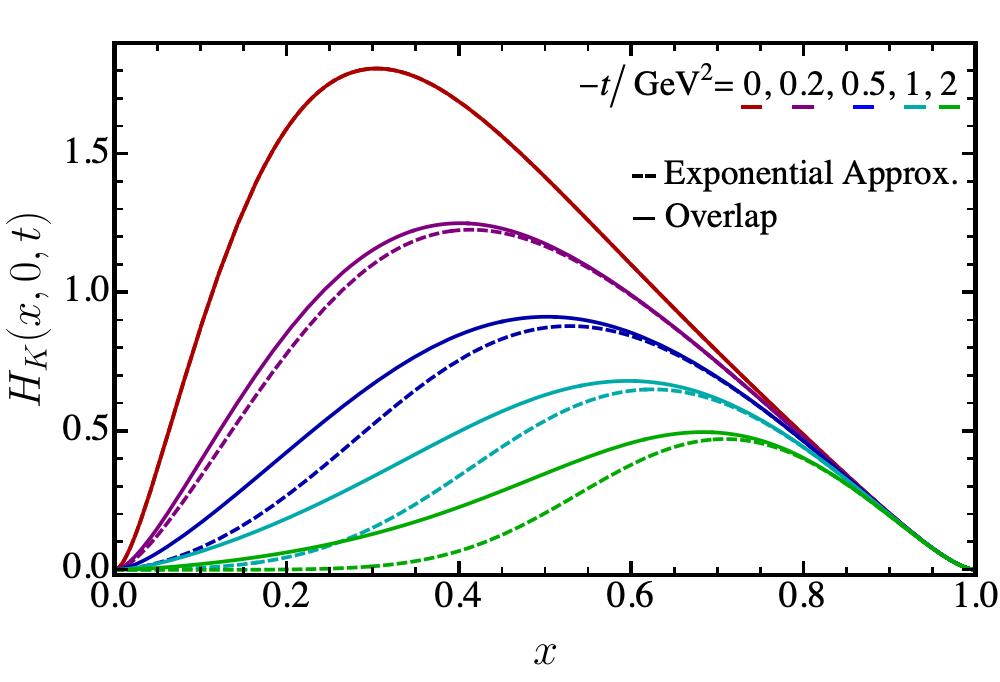}
    \includegraphics[scale=.24]{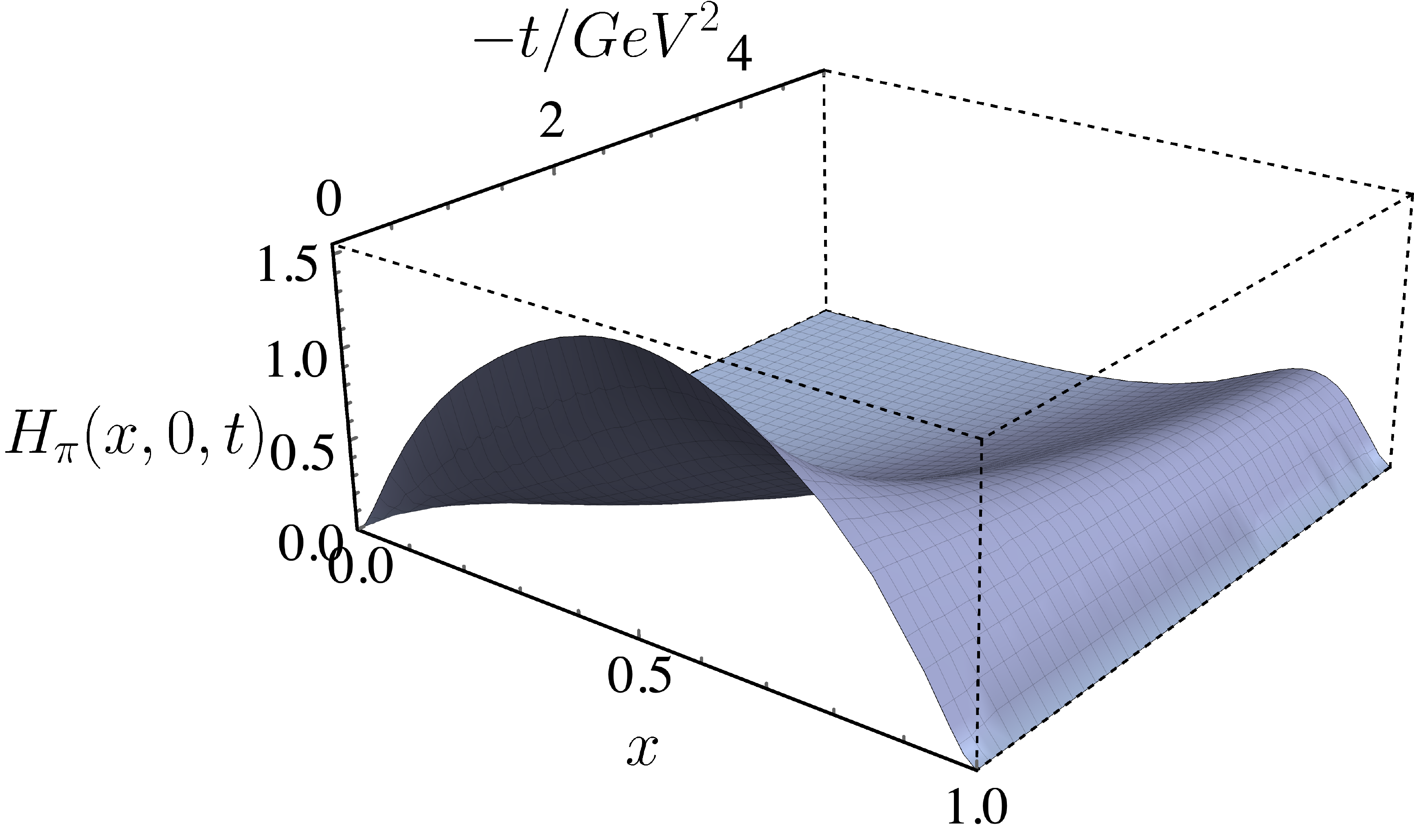}
    \includegraphics[scale=.24]{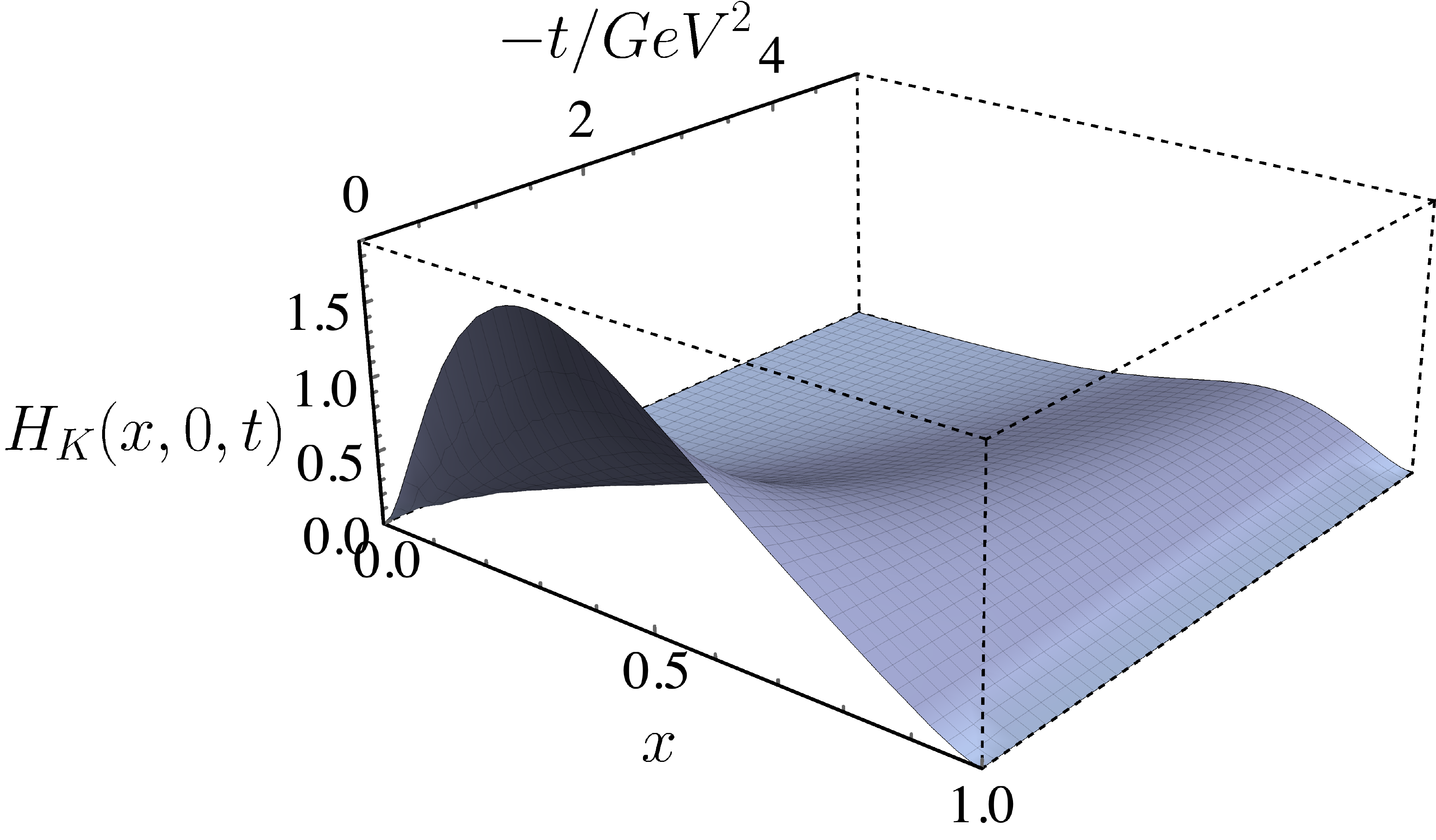}
    \caption{Zero-skewness valence quark GPDs for pion (left) and kaon (right), employing the model inputs described in Section V. Upper panel- A comparison between the GPDs obtained directly from Eq.~(\ref{eq:GPDfinal}) [solid lines] and those produced by the algebraic representation in Eq.~\eqref{eq:GPDLFHQCD} [dashed lines]. Lower panel- Equivalent three-dimensional picture, resulting from Eq.~(\ref{eq:GPDfinal}). Mass units in GeV.}
    \label{fig:Pion-Kaon-GPDs}
\end{figure*}
\section{Generalized parton distributions}
The valence quark GPD can be obtained from the overlap representation of the LFWF~\cite{Diehl:2003ny}, namely:
\begin{eqnarray}
\nonumber
 H_{\M}^q(x,\xi,t) \hspace{-0.1 cm} &=& \hspace{-0.1 cm} \int \frac{d^2k_{\perp}}{16\pi^3} \psi_{\M}^{q*}  \left( x^-, (\mathbf{k}_\perp^{-})^2\right) \psi_{\M}^q  \left( x^+, (\mathbf{k}^{+}_\perp)^2 \right) \,, \\
\hspace{0.2 cm} x^{\pm}&=&\frac{x\pm \xi}{1\pm \xi}\hspace{0.4 cm} , \hspace{0.4 cm} \mathbf{k}_\perp^\pm=k_\perp \mp \frac{\Delta_\perp}{2}\frac{1-x}{1\pm \xi}\;.
\label{GPDdefinition}
\end{eqnarray}
If $p\,(p')$ denotes the initial (final) meson momentum, then $P=(p+p')/2$ and $-t=\Delta^2=(p-p')^2$ (the latter defines the momentum transfer); $\Delta_\perp^2 = \Delta^2(1-\xi^2) -4\xi^2 m_{\text{M}}^2$. In addition, the longitudinal momentum fraction transfer is  $\xi=[-n\cdot \Delta]/[2n\cdot P]$.  Both $x$ and $\xi$ have support on $[-1,1]$, but the overlap representation is only valid in the DGLAP region, $|x| > \xi$. The kinematical completion (the extension to the ERBL domain), required to fulfill the polinomiality property~\cite{Diehl:2003ny}, can be achieved through the covariant extension from Refs.~\cite{Chouika:2017dhe,Chouika:2017rzs,Chavez:2021koz,Chavez:2021llq}. Notwithstanding, the GPD is even in $\xi$ and only non-zero for the valence quark if $x > -\xi$ (the antiquark GPD is non-zero if $x < \xi$); hence, in the following, we shall restrain ourselves to $\xi \geq 0$. Notice again that Eq.~\eqref{GPDdefinition} implies that the meson is described as a two-body Fock state. This picture is then valid at the hadronic scale, in which the fully dressed quark/antiquark quasiparticles encode all the properties of the meson. 

We now work out the expression for the valence quark GPD in detail by substituting Eq.~\eqref{eq:LFWFPDArel} in Eq.~\eqref{GPDdefinition}
\begin{eqnarray}
\nonumber
H_{\M}^q(x,\xi,t)=(16\pi^2 f_{\M}\nu)^2  \phi_{\M}^q(x^+)\phi_{\M}^q(x^-) \Lambda_{1-2x^+}^{2\nu} \Lambda_{1-2x^-}^{2\nu} \\
\times
\int \frac{d^2k_\perp}{16\pi^3} \frac{1}{((\bold{k}_\perp^-)^2+\Lambda_{1-2x^-}^2)^{\nu+1}}\frac{1}{((\bold{k}_\perp^+)^2+\Lambda_{1-2x^+}^2)^{\nu+1}} \;. \nonumber \\
\label{eq:GPDint1}
\end{eqnarray}
As usual, integration on $k_\perp$ can be performed by introducing Feynman parametrization and a suitable change of variables, such that the integral in Eq.~\eqref{eq:GPDint1} becomes
\begin{eqnarray}\nonumber
\frac{2\pi}{16\pi^3} \frac{\Gamma(2\nu+2)}{\Gamma^2(\nu+1)} && \hspace{-0.02 cm} \int_0^1 \hspace{-0.1 cm} du u^\nu(1-u)^\nu \hspace{-0.1 cm} \int_0^\infty \hspace{-0.2 cm} dk_\perp \frac{k_\perp}{(k_\perp^2+\mathbb{M}^2(u))^{2\nu+2}} \\
 && \hspace{-0.5cm} = \frac{1}{16\pi^2}\frac{\Gamma(2\nu+1)}{\Gamma^2(\nu+1)}\int_0^1 du \frac{u^\nu (1-u)^\nu}{[\mathbb{M}^2(u)]^{2\nu+1}} \;,
\label{eq:GPDint2}
\end{eqnarray}
\begin{figure*}[!ht]
    \centering
    \includegraphics[scale=.24]{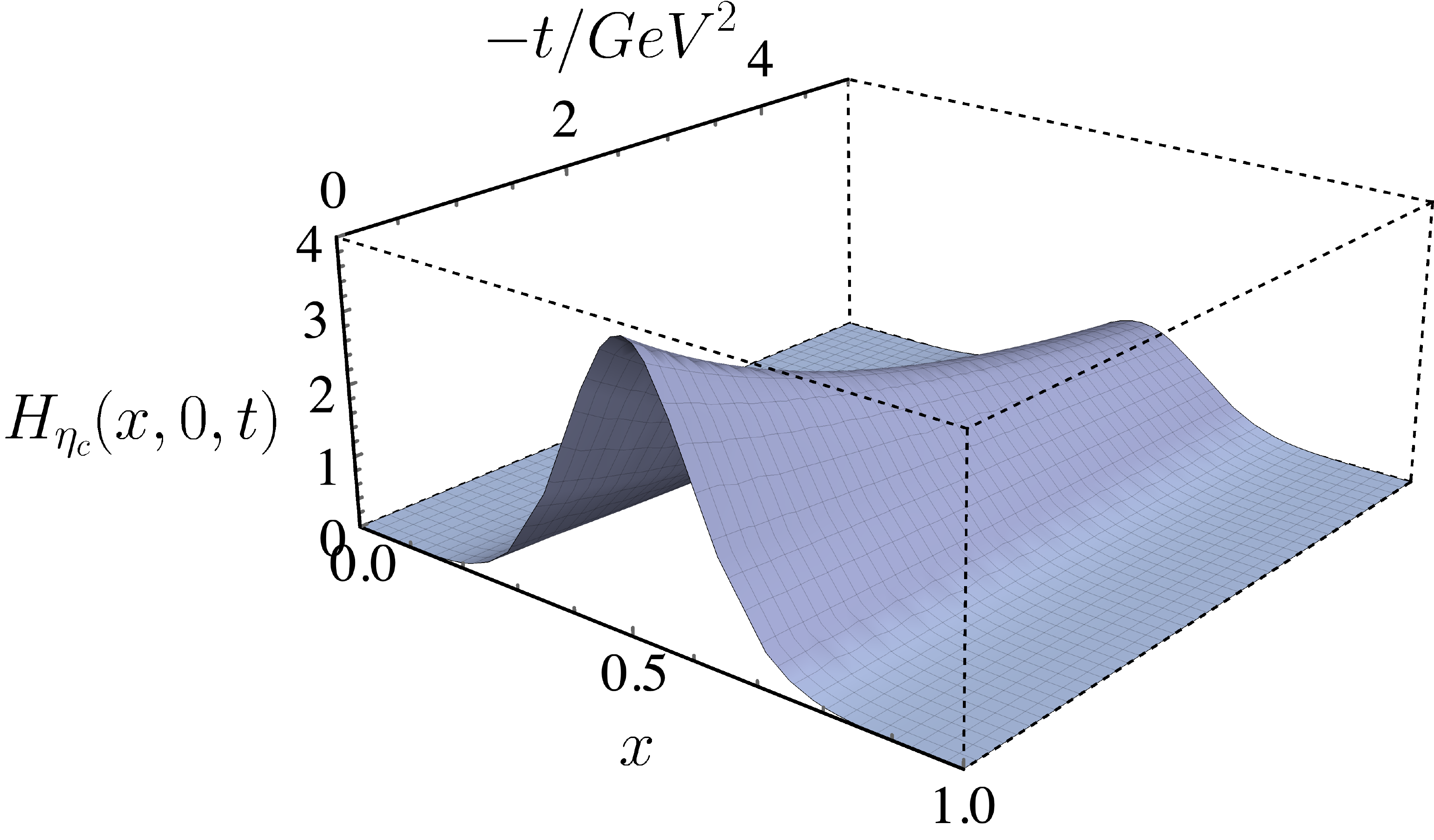}
    \includegraphics[scale=.24]{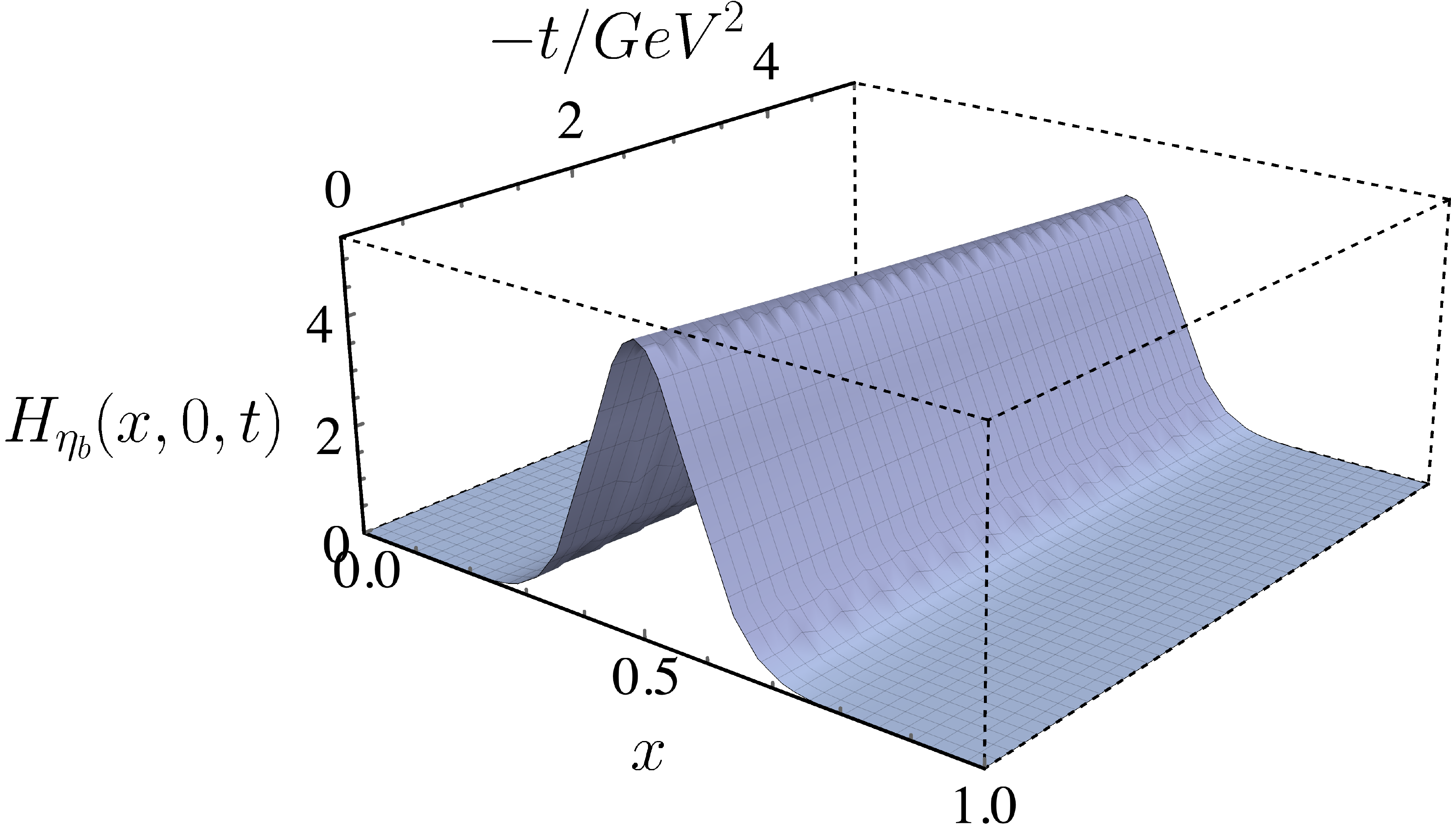}
    \caption{Valence quark GPDs obtained from Eq.~(\ref{eq:GPDfinal}) for $\xi=0$ employing the model inputs described in Section V. Left panel- $\eta_c$ GPD. Right panel- $\eta_b$ GPD. Mass units in GeV.}
    \label{EtacGPD}
\end{figure*}
where the function $\mathbb{M}^2(u)$ depends on the model parameters, as well as the kinematic variables $x,\;\xi,\;t$. It acquires the form $\mathbb{M}^2(u)=c_2 u^2 + c_1 u + c_0$, where
\begin{eqnarray}
\nonumber
c_2 &=& \frac{(1-x)^2}{(1-\xi^2)^2}t \;, \\
\nonumber
c_1 &=& -\frac{(1-x)^2}{(1-\xi^2)^2}t + \Lambda^2_{1-2x^+}-\Lambda^2_{1-2x^-} \;, \\
c_0 &=& \Lambda^2_{1-2x^-}\;.
\end{eqnarray}
Thus the GPD can be conveniently expressed as
\begin{eqnarray}
\nonumber
H_{\M}^q(x,\xi,t)&=&\mathcal{N} \phi_{\M}^q(x^+)\phi_{\M}^q(x^-) \Lambda_{1-2x^+}^{2\nu} \Lambda_{1-2x^-}^{2\nu}   \\
&\times& \frac{\Gamma(2\nu+2)}{\Gamma^2(\nu+1)}\int_0^1 du \frac{u^\nu (1-u)^\nu}{[\mathbb{M}^2(u)]^{2\nu+1}}\;.
\label{eq:GPDfinal}
\end{eqnarray}
Notice that, in the chiral limit, $\mathbb{M}^2(u)$ reduces to
\begin{equation}
    \mathbb{M}^2(u)= -t\,u(1-u)\frac{(1-x)^2}{(1-\xi^2)^2}+M_q^2\;,
\end{equation}
and so the integration on $du$ in Eq.~\eqref{eq:GPDfinal} can be carried out algebraically for specific values of $\nu\,\textgreater -1$. In particular, $\nu = 1$ recovers the results in~\cite{Chouika:2017rzs, Mezrag:2016hnp, Mezrag:2014jka,Chavez:2021koz,Chavez:2021llq}. Beyond the chiral limit, an algebraic expression is found for $t = 0$:
\begin{eqnarray}
H_{\M}^q(x,\xi,0)&=&\mathcal{N} \phi_{\M}^q(x^+)\phi_{\M}^q(x^-) \frac{\Lambda_{1-2x^+}^{2\nu}}{\Lambda_{1-2x^-}^{2\nu}}  \frac{\Gamma(2\nu+2)}{\Lambda_{1-2x^-}^2} \\
&& \hspace{-0.3 cm} \times \;_2\tilde{F}_1\left(1+\nu,1+2\nu,2\nu+2,1-\frac{\Lambda_{1-2x^+}^{2}}{\Lambda_{1-2x^-}^{2}}\right) \,,\nonumber
\label{eq:GPDxi}
\end{eqnarray}
where $_p\tilde{F}_q(u,v,w,z)$ is the regularized hypergeometric function. Conversely, taking $\xi=0$, an expansion of $\mathbb{M}^2(u)$ around $-t\approx 0$ yields an algebraic solution for Eq.~\eqref{eq:GPDfinal}:
\begin{eqnarray}
\label{eq:exptGPD}
H_{\M}^q(x,0,t) \hspace{-0.1 cm} &\overset{t\to 0}{\approx}& \hspace{-0.1 cm} \mathcal{N}\frac{[\phi_{\M}^{q}(x)]^2}{\Lambda_{1-2x}^2} \hspace{-0.1 cm} \left[1-c_\nu^{(1)}(1-x)^2 \hspace{-0.1 cm} \left( \hspace{-1mm} \frac{-t}{\Lambda_{1-2x}^2} \hspace{-1mm} \right) \hspace{-0.1 cm} + ... \right] \hspace{-0.1 cm} \,, \nonumber \\
&& \hspace{-2cm} c_\nu^{(1)}= \frac{(1+\nu)(1+2\nu)}{2(3+2\nu)}\;,\;\mathcal{N}=\left[\int_0^1 dx\; \frac{\phi_{\M}^2(x)}{\Lambda_{1-2x}^2}  \right]^{-1}\;.
\end{eqnarray}
In the next section we will focus on the \emph{forward limit} of the GPD ($t=0$, $\xi=0$) which defines the valence quark PDF. For the time being, we can make an insightful connection with light-front holographic QCD (LFHQCD) approach Ref.~\cite{Chang:2020kjj,deTeramond:2018ecg}, recalling the following representation for the zero-skewness valence quark GPD therein:
\begin{equation}
\label{eq:GPDLFHQCD}
    H_{\M}^q(x,0,t)= q_{\M}(x) \;\text{exp}[t \hat{f}_{\M}^q(x)]\;,
\end{equation}
where $\hat{f}_{\M}^q$ is some profile function to be determined. An expansion around $-t \approx 0$ of this expression, and a subsequent comparison with Eq.~\eqref{eq:exptGPD}, enable us to identify 
\begin{equation}
    \label{eq:fforGPDaprox}
    \hat{f}_{\M}^q(x) = \frac{c_\nu^{(1)}(1-x)^2}{\Lambda_{1-2x}^2} \;.
\end{equation}
The parametric representation of the GPD in Eq.~\eqref{eq:GPDLFHQCD} provides a fair approximation of the zero-skewness GPD in Eq.~\eqref{eq:GPDfinal} except for intermediate values of momentum transfer. It is also useful in extracting insights concerning the IPS-GPDs, as will be addressed below.

We now proceed to discuss the derivation of PDFs, FFs and IPS-GPDs, as inferred from the knowledge of the GPDs in the DGLAP kinematic region. 
\begin{figure}[!ht]
    \centering
    \includegraphics[scale=.24]{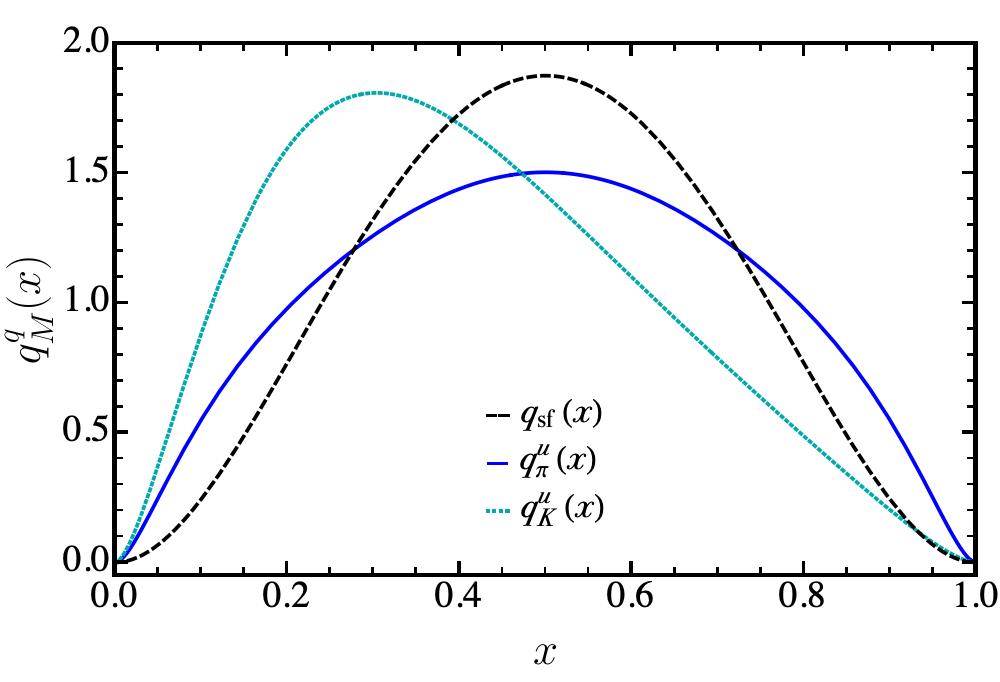}
    \includegraphics[scale=.24]{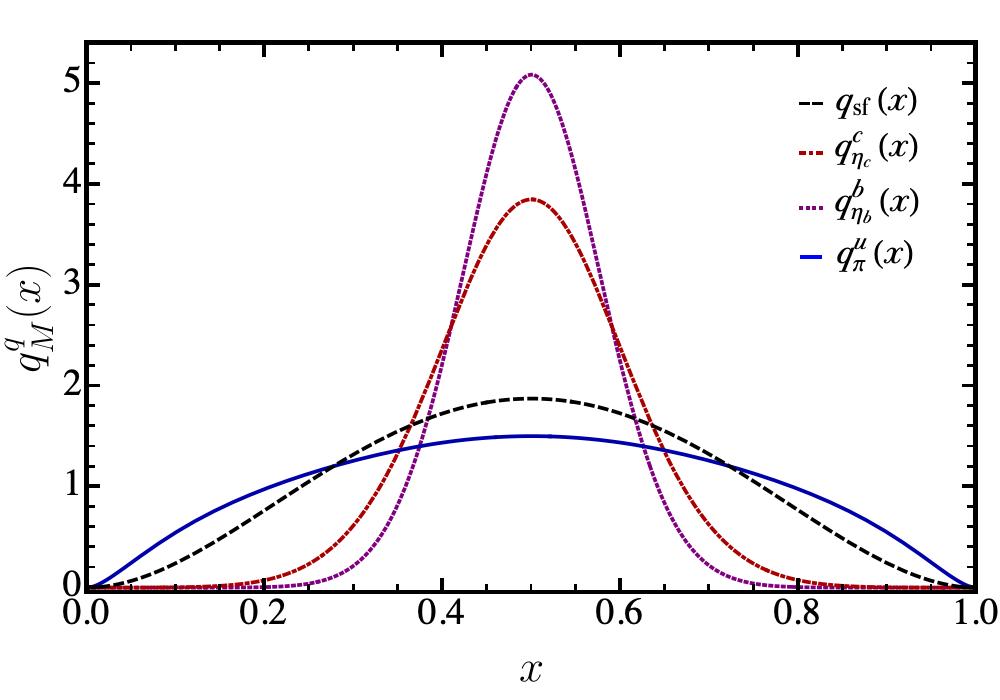}
    \caption{Valence quark PDFs at $\zeta_H$. Upper panel- The solid (blue) line corresponds to pion and the dotted (cyan) line corresponds to the light-quark PDF in kaon. 
    Lower panel- The dot-dashed (red) line corresponds to $\eta_c$, the dotted (purple) line corresponds to $\eta_b$, again the solid (blue) line corresponds to pion. For all these panels, the dashed (black) line corresponds to the parton like profile $q_{sf}(x)=30x^2(1-x)^2$.}
    \label{PDFs}
\end{figure}

\subsection{Parton distribution functions}

The first term of the Taylor expansion in Eq.~\eqref{eq:exptGPD} corresponds to the valence quark PDF, namely
\begin{eqnarray}
\label{eq:defPDF}
    q_{\M}(x)&\equiv& H_{\M}^q(x,0,0) = \mathcal{N} \frac{[\phi_{\M}^{q}(x)] ^2}{\Lambda^2_{1-2x}}\;,
\end{eqnarray}
where $q_{\M}(x)$ is unit normalized. Recalling that the distributions have been derived at $\zeta_H$, the corresponding antiquark PDF is simply obtained as
\begin{equation}
\label{eq:antiPDF1}
    \bar{h}_{\M}(x;\zeta_H) = q_{\M}(1-x;\zeta_H)\;.
    \end{equation}
Furthermore, the factorization properties of the LFWF in the chiral limit yields the simple relation:
\begin{equation}
\label{fac:PDF}
    q_{\M}(x;\zeta_H)=\frac{[\phi_{\M}^{q}(x;\zeta_H)]^2}{\int_0^1 dx\; [\phi_{\M}^{q}(x;\zeta_H)]^2},
\end{equation}
\begin{figure}[!ht]
    \centering
    \includegraphics[scale=.24]{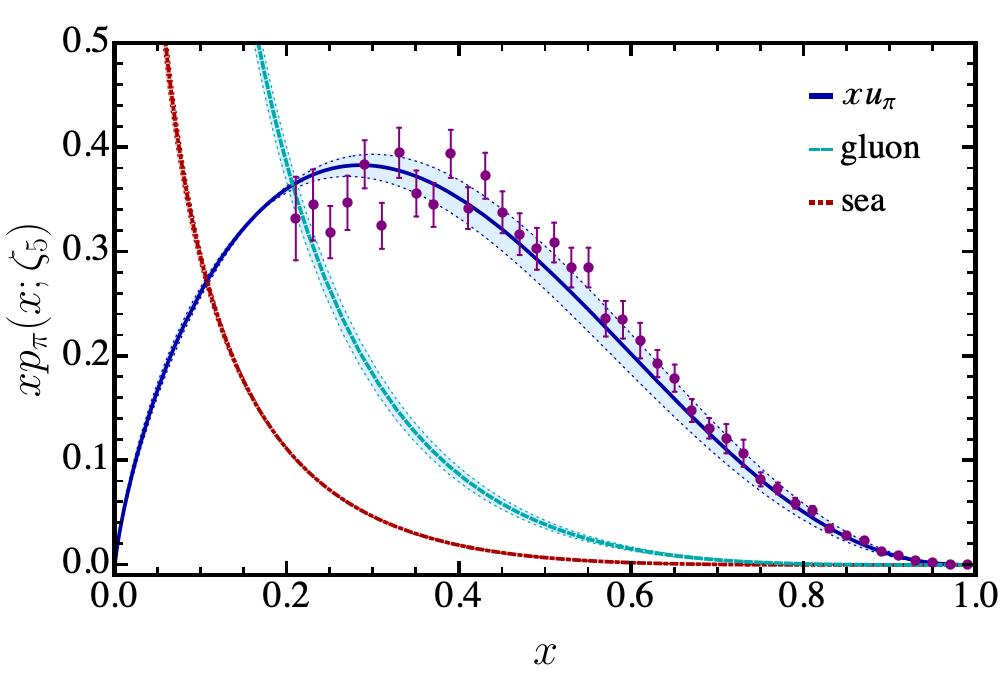}
    \includegraphics[scale=.24]{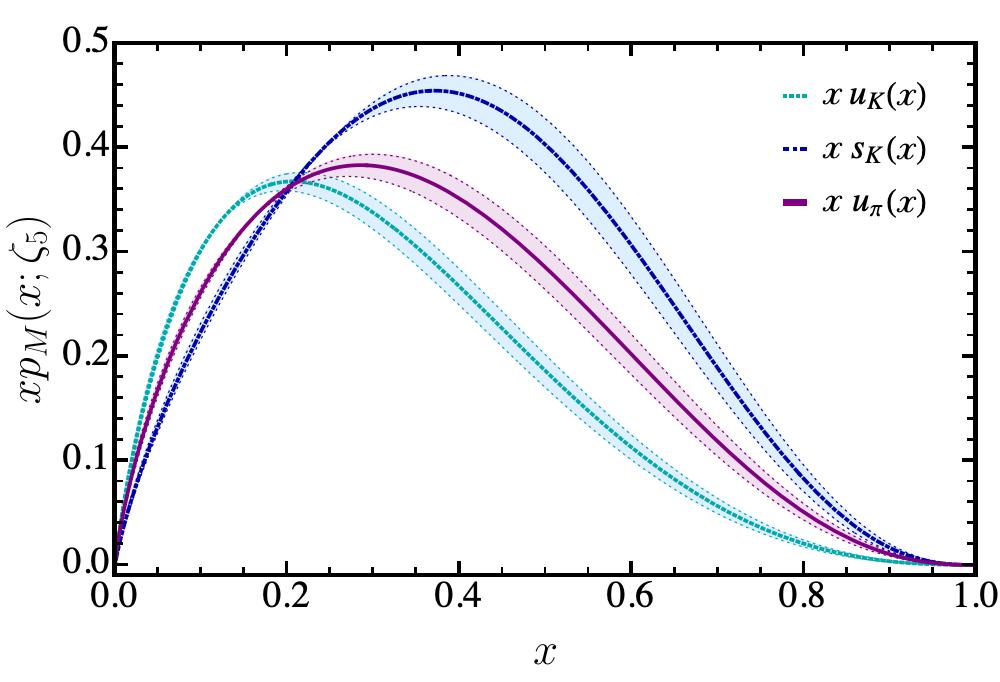}
    \caption{Evolved PDFs at $\zeta_5:=5.2$ GeV. Upper panel- The plots correspond to the evolved pion PDF. The solid (blue) line corresponds to the $u$ valence-quark, the dashed (cyan) line corresponds to the gluon contribution and the dot-dashed (red) line corresponds to sea contribution. The data from~\cite{PhysRevD.39.92}, is rescaled according to the ASV analysis in~\cite{PhysRevLett.105.252003}. Lower panel- The dotted (cyan) line corresponds to $u$-in-$K$ valence-quark PDF, the dot-dashed (blue) line is the analogous for the $\bar{s}$ quark  and the solid (purple) line corresponds to the $u$ valence-quark in the pion. The error bands account for the variation of the initial scale, $\zeta_H=0.33\,(1\pm0.1)$ GeV.}
    \label{PDFsE}
\end{figure}
thus stressing that the degree of factorizability of the AM is manifest via the quantity $\Lambda^2_{1-2x}$. As long as we have $m_{\M}^2\approx 0\;\text{and also}\;(M_{\bar{h}}^2-M_q^2) \approx 0$, a factorized LFWF will produce sensible results. This is the case of the SDE results from Refs.~\cite{Cui:2020dlm,Cui:2020tdf}, in which Eq.~\eqref{fac:PDF} was employed to compute the kaon PDF from its PDA. For the purpose of this work, factorizability will not be assumed and we shall consider the more general case, Eq.~\eqref{eq:defPDF}. The set of relations described in this Section also shows that if the input PDA behaves like $\phi(x\to1) \sim (1-x)$ (as prescribed by QCD,~\cite{Lepage:1979zb}), the PDF will exhibit the large-$x$ behavior $q_{\M}(x; \zeta_H) \sim (1-x)^2$. Finally, it is worth recalling that neither Eq.~\eqref{fac:PDF} nor Eq.~\eqref{eq:antiPDF1} remain valid for $\zeta \textgreater \zeta_H$, due to the evolution equations obeyed by the PDFs~\cite{Dokshitzer:1977sg, Gribov:1972ri, Lipatov:1974qm, Altarelli:1977zs}.

All distributions described so far have been obtained from the LFWF at the hadron scale, $\zeta_H$; as described before, at this low-energy scale, the fully dressed quasiparticles (valence-quarks) express all hadron properties. This is also the case of the valence-quark PDF which, computed at $\zeta_H$, entails that all the hadron's momentum is carried by the fully-dressed valence quarks. From the experimental point of view, the access and interpretation of PDFs and GPDs at $\zeta_H$ imply certain technical and conceptual complications~\cite{Ellis:1996mzs}; only above certain energies, typically the mass of the proton, parton distributions can be properly extracted. In particular, experimental data for the case of the pion is only available at $\zeta=\zeta_5:=5.2$ GeV~\cite{PhysRevLett.105.252003,Conway:1989fs} (the same for the $u_K(x)/u_\pi(x)$ ratio~\cite{Orsay:1980fhh}), whereas $\zeta=\zeta_2:=2$ GeV is a typical scale for lattice QCD and phenomenological fits~\cite{Sufian:2020vzb,Joo:2019bzr,Sufian:2019bol}. To produce a consistent picture when evolving the hadronic scale PDF, we shall follow the all orders scheme introduced in Refs.~\cite{Rodriguez-Quintero:2019fyc,Ding:2019lwe,Ding:2019qlr,Cui:2020tdf,Cui:2020dlm} for pion and kaon PDFs, extended to their GPDs in Ref.~\cite{Raya:2022eqa,Raya:2021zrz}, and employed recently in the calculation of the proton PDFs as well~\cite{Lu:2022cjx}. This scheme  is based upon the assumption that an effective charge $\hat{\alpha}$ allows all beyond leading-order effects to be absorbed within it, thus arriving at a leading-order-like DGLAP evolution equation. Notably, if the evolution is performed via the computation of several Mellin moments, it is not necessary to specify the pointwise behavior of the effective charge~\cite{Raya:2021zrz} (assuming its existence would be sufficient). To evolve the distributions directly, the exercise we carry out in this article, we take $\hat{\alpha}$ from Ref.~\cite{Cui:2020tdf}, which implies setting $\zeta_H=0.33(1\pm0.1)$ GeV. In Section V, we present numerical results for evolved pion and kaon PDFs for specific model inputs described therein.
\begin{figure*}[!htb]
    \centering
    \includegraphics[scale=.24]{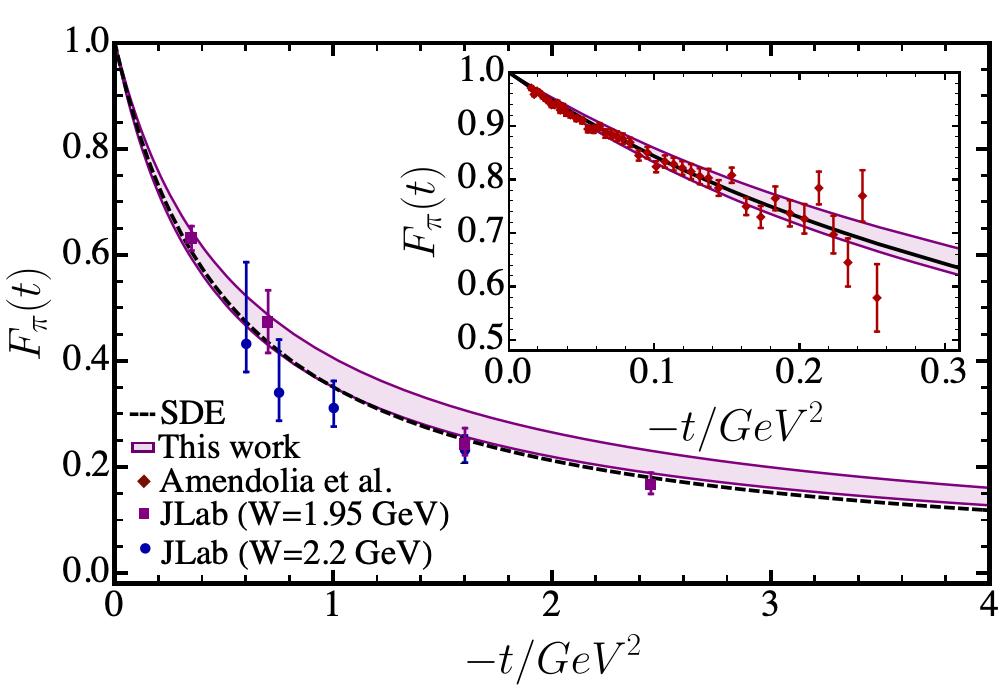}
    \includegraphics[scale=.24]{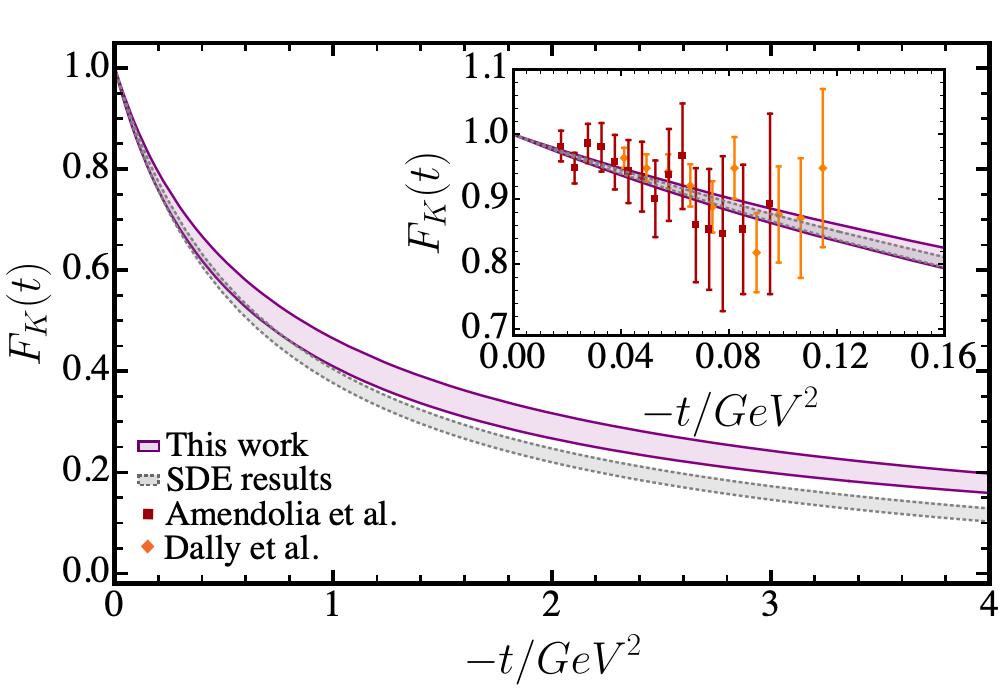}
    \caption{Pion and kaon electromagnetic FFs. Left panel- The (purple) band represents our pion results with the model parameters described in Section V. The band width accounts for a $5\%$ variation of the benchmark charge radius in Table 1. Dashed (black) line is the SDE result for the pion~\cite{Chang:2013nia}. Right panel- It shows analogous results for the kaon FF. Diamonds, rectangles and circles represent the experimental data from Refs.~\cite{AMENDOLIA1986168,JeffersonLabFpi:2000nlc,JeffersonLabFpi-2:2006ysh}. Lower (gray) band is the SDE result for the kaon~\cite{Eichmann:2019tjk}.}
    \label{Pion-Kaon-FF}
\end{figure*}
\subsection{Electromagnetic form factor}
\begin{figure*}[!htb]
    \centering
    \includegraphics[scale=.24]{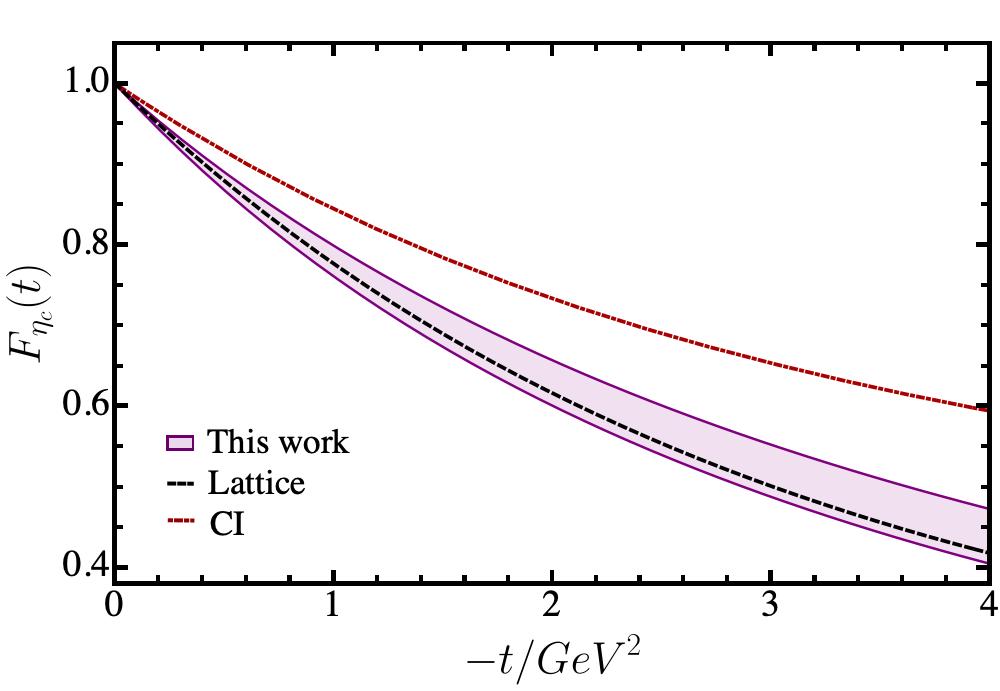}
    \includegraphics[scale=.24]{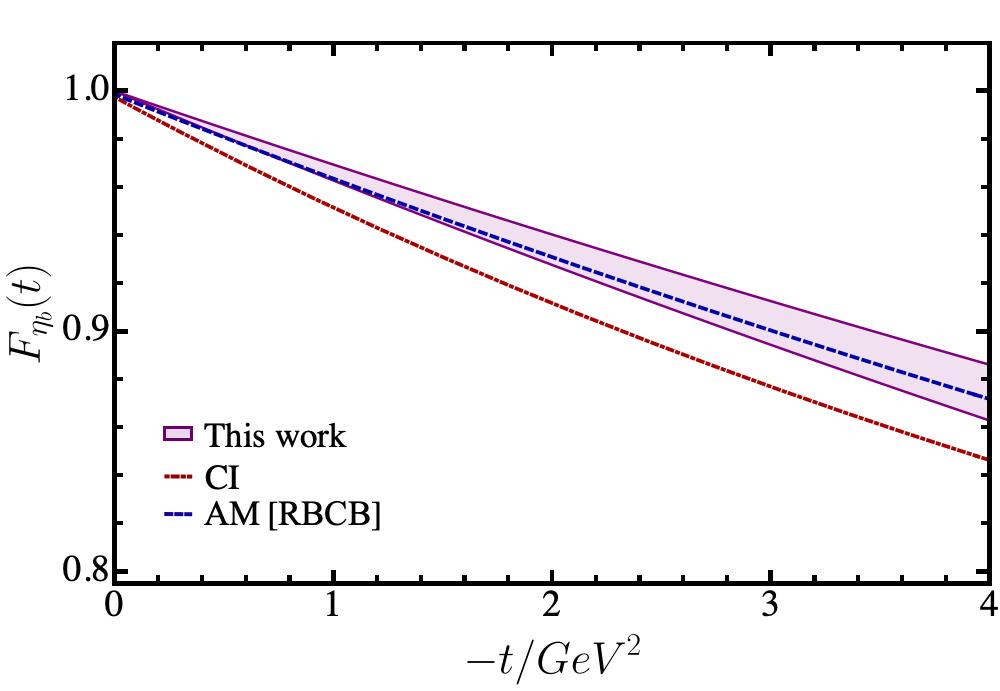}
    \caption{$\eta_c$ and $\eta_b$ electromagnetic FFs. Left panel- The purple band represents our $\eta_c$ results with the model parameters described in Section V. The band width accounts for a $5\%$ variation of the benchmark charge radius in Table 1. Right panel- Analogous results for $\eta_b$. For comparison, we have included lattice QCD results from Refs.~\cite{Dudek:2006ej,Dudek:2007zz}, as well as SDE-driven predictions in the contact interaction (CI) model and a former algebraic model for heavy quarkonia~\cite{Bedolla:2016yxq,Raya:2017ggu}.}
    \label{EtacFF}
\end{figure*}

The contribution of the $q$ quark to the meson's elastic electromagnetic form factor (EFF) is obtained from the zeroth moment of the GPD:
\begin{equation}
\label{eq:EFFq}
    F_{\M}^q(t) = \int_{-1}^1 dx\; H_{\M}^q(x,\xi,t),
\end{equation}
an analogous expression holds for the antiquark $\bar{h}$, such that the complete meson EFF reads
\begin{equation}
    F_{\M}(t) = e_q F_{\M}^q(t)+e_{\bar{h}}F_{\M}^{\bar{h}}(t)\;, \label{eq:MFF1}
\end{equation}
where $e_{q,\bar{h}}$ are the valence-constituent quarks electric charges in units of the positron charge. Due to polynomiality properties of the GPD, the EFF does not depend on $\xi$, therefore one can simply take $\xi \to 0$:
\begin{equation}
\label{eq:EFFq2}
    F_{\M}^q(t) = \int_{0}^1 dx\; H_{\M}^q(x,0,t)\;.
\end{equation}
A Taylor expansion around $t\approx 0$ yields
\begin{eqnarray}
\label{eq:chargeradius}
F_{\M}^q(t) &\overset{t\to 0}{\approx}&1-\frac{(r_{\M}^q)^2}{6}(-t)+...\;,\\
(r_{\M}^q)^2 &=&\left. -6 \frac{d F_{\M}^q(t)}{dt} \right|_{t=0}\;,
\end{eqnarray}
where $r_{\M}^q$ denotes the contribution of the quark $q$ to the meson charge radius, $r_{\M}$. Comparing the above equations with the integration of Eq.~\eqref{eq:exptGPD}  on $x$, one obtains a semi-analytical expression for $r_{\M}^q$:
\begin{equation}
\label{eq:crquark}
    (r_{\M}^q)^2=6 \, \int_0^1 dx\;\hat{f}_{\M}^q(x)q_{\M}(x)\;,
\end{equation}
showing the charge radius is tightly connected with the hadronic scale PDF (and thus with the corresponding PDA). The antiquark result is obtained analogously. This contribution to $r_{\M}$ reads:
\begin{equation}
\label{eq:crantiquark}
    (r_{\M}^{\bar{h}})^2=6 \,\int_0^1 dx\;\hat{f}_{\M}^{\bar{h}}(x)q_{\M}(1-x)\;,
\end{equation}
where $\hat{f}_{\M}^{\bar{h}}(x)$ is defined in analogy to its quark counterpart in Eq.~\eqref{eq:fforGPDaprox},
\begin{equation}
    \hat{f}_{\M}^{\bar{h}}(x) = \frac{c_\nu^{(1)}(1-x)^2}{\Lambda_{2x-1}^2}\;.
\end{equation}
Summing up the quark and antiquark contributions, the meson charge radius reads:
\begin{equation}
\label{eq:crgeneral}
    r_{\M}^2= e_q (r_{\M}^q)^2 + e_{\bar{h}} (r_{\M}^{\bar{h}})^2\;.
\end{equation}
Clearly, in the isospin symmetric limit, $e_q+e_{\bar{h}}=1$ yields  $F_{\M}(t) = F_{\M}^q(t)$ and so $r_{\M} = r_{\M}^q$ . For the neutral pseudo-scalars, in the isospin symmetric limit, $e_q+e_{\bar{h}}=0$ implies $F_{\M}$ would be strictly zero, producing $r_{\M}=0$; thereby we focus only on the individual flavor contribution ($F_{\M} \to F_{\M}^q$) in such cases (\emph{e.g.} heavy quarkonia). Finally, note that if the charge radius is known, then Eqs.~(\ref{eq:crquark}- \ref{eq:crgeneral}) can be employed to fix the model parameters.

\subsection{Impact parameter space GPD}

\begin{figure*}[!htb]
    \centering
    \includegraphics[scale=.4]{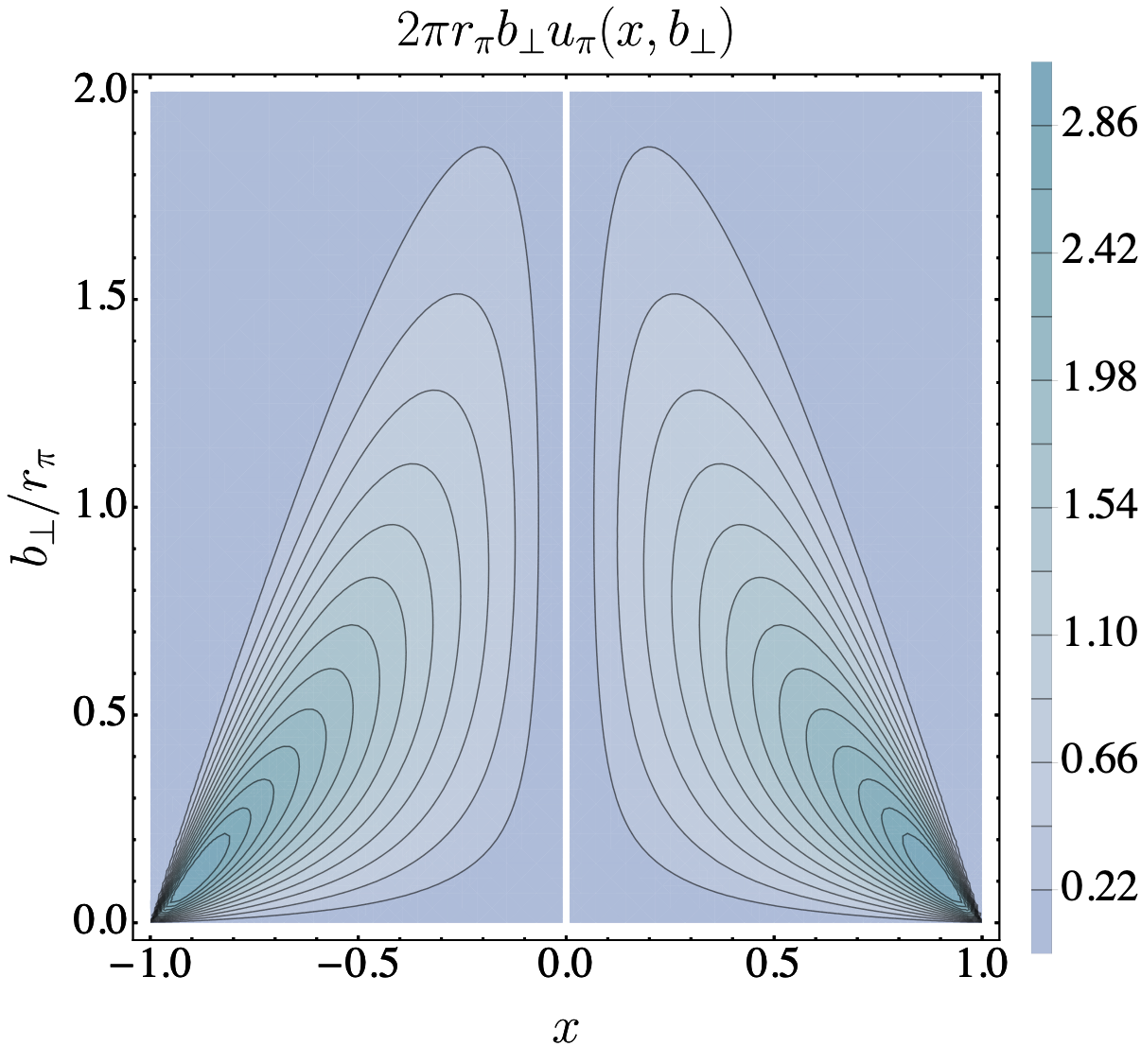}
    \includegraphics[scale=.4]{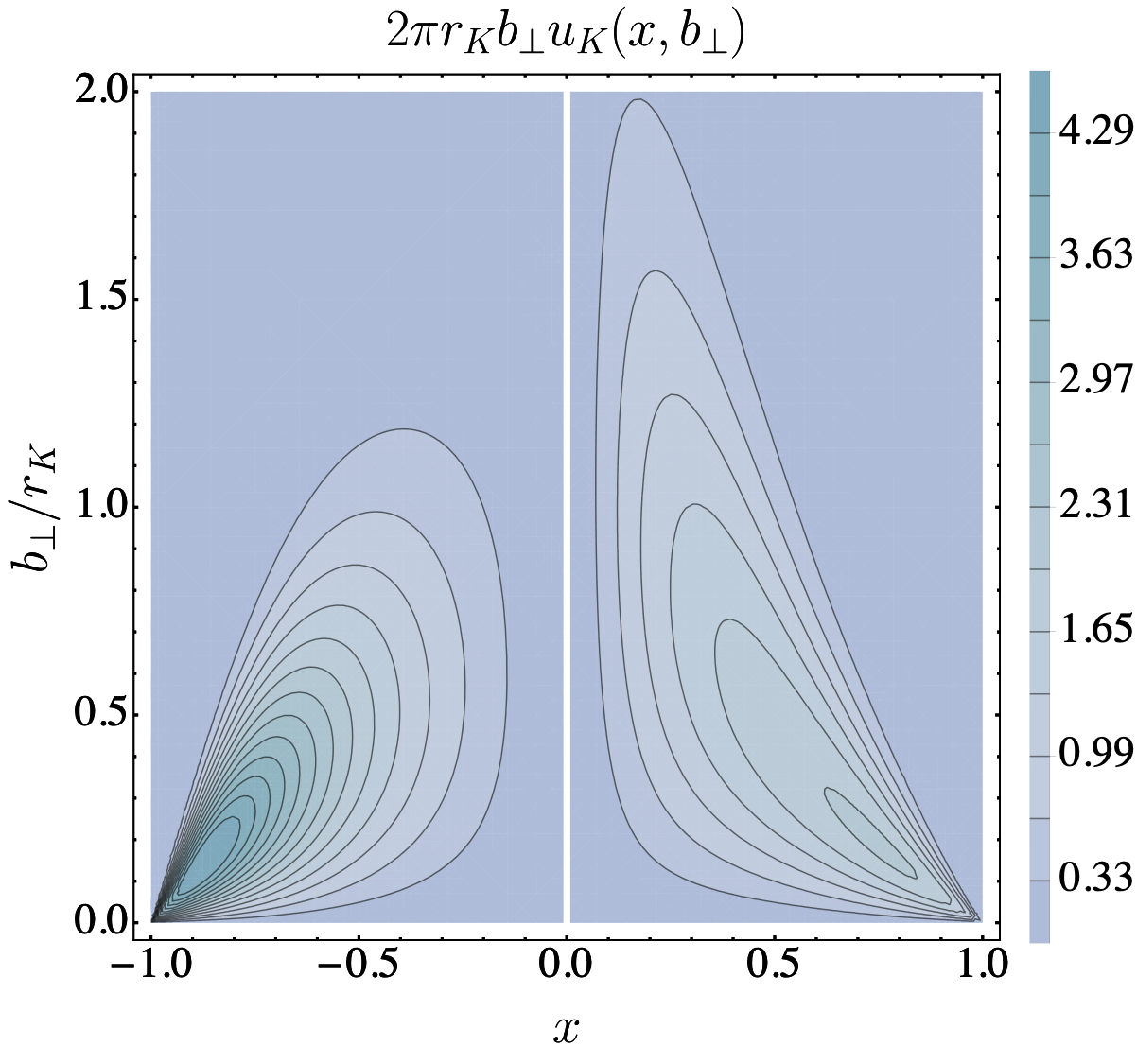}
    \caption{Impact parameter space GPDs. The quark lies in the $x\textgreater 0$ domain, while the antiquark in $x\textless 0$. Left panel- pion results using the inputs from Section V. Right panel- analogous results for the kaon. The conspicuous asymmetry in this case is due to the larger $s$-quark mass: the $s$-quark plays a larger role in determining the center of transverse momentum.   
    }
    \label{PionIP}
\end{figure*}
The IPS-GPD can be obtained straightforwardly by carrying out the Fourier transform of the zero-skewness GPD, $H_{\M}^q(x,0,t)$:
\begin{eqnarray}
u_{\M}(x,b_{\perp}^2)= \int_0^{\infty}\frac{d\Delta}{2\pi }\Delta J_0 (b_{\perp} \Delta) H_{\M}^u(x,0,t),
\end{eqnarray} 
where $J_0$ is the cylindrical Bessel function. This distribution is interpreted as 
the probability density of finding a parton with momentum fraction $x$ at a transverse distance $b_{\perp}$ from the centre of transverse momentum of the meson under study. It is extracted in its totality
by the GPD’s properties in the DGLAP region. Exploiting the representation of the GPD from
Eq.~(\ref{eq:GPDLFHQCD}), we can obtain an analytic expression:
\begin{eqnarray}
\label{eq:IPSux}
u_{\M}^q(x,b_{\perp}^2)= \frac{q_{\M}(x)}{4 \pi \hat{f}_{\M}^{q}(x)} \text{exp}\left[-\frac{b_\perp^2}{4\hat{f}_{\M}^{q}(x)}\right]\;.
\end{eqnarray}
Containing an explicit dependence on the PDF, Eq.~\eqref{eq:IPSux} reveals a clear interrelation between the momentum and spatial distributions. In fact, the PDF is recovered from
\begin{equation}
    q_{\M}(x)=2\pi \int_0^\infty db_\perp\, b_\perp q(x,b_\perp)\;.
\end{equation}
Furthermore, considering the mean-squared transverse extent (MSTE),
\begin{eqnarray}
    \textless  b_\perp^2(x) \textgreater_{\M}^q &=& \frac{1}{r_{\M}}\int_0^\infty db_\perp\, \textbf{b}_{\M}^q(x,b_\perp) \,b_\perp^2 \;,\\
    \textbf{b}_{\M}^q(x,b_\perp) &:=& 2\pi r_{\M} b_\perp u_{\M}^q(x,b_\perp)\;.\label{eq:bqblack}
\end{eqnarray}
the IPS-GPD defined in Eq.~\eqref{eq:IPSux} yields the plain relation:
\begin{equation}
    \textless  b_\perp^2(x) \textgreater_{\M}^q = 4\int_0^1 dx\,\hat{f}_{\M}^{q}(x)q_{\M}(x)\;.
\end{equation}
Integrating over $x$, and comparing with Eq.~\eqref{eq:crgeneral}, one is left with a compact expression for the expectation value:
\begin{equation}
    \textless b_\perp^2 \textgreater_{\M}^q =  \frac{2}{3}r_{\M}^2  \left[ \frac{(r_{\M}^q)^2}{e_q(r_{\M}^q)^2+e_{\bar{h}}(r_{\M}^{\bar{h}})^2} \right]\;;\label{eq:aveMSTE}
\end{equation}
\emph{i.e.} the expectation value of the MSTE of the valence quark is directly correlated with the meson charge radius. In the isospin symmetric limit, the following expected result~\cite{Raya:2022eqa,Raya:2021zrz} is recovered:
\begin{equation}
\label{eq:expMSTE}
    \textless b_\perp^2 \textgreater_{\M}^q =  \frac{2}{3}r_{\M}^2\;.
\end{equation}
Interestingly, in the chiral limit, all the algebraic expressions form this Section, valid only at $\zeta_H$, become plainly analogous to those from the factorized \emph{Gaussian model} in~\cite{Raya:2022eqa,Raya:2021zrz}.

In the following section we shall provide a collection of results for the distributions discussed so far, using SDE predictions as model inputs.

\section{Computed distributions}
Now that we have shown a variety of algebraic results for different distributions of partons (and some other quantities), we will particularize the inputs of the AM. The starting point is Eq.~\eqref{eq:LFWFPDArel}, which directly relates the leading-twist LFWF with the PDA such that, with the prior knowledge of $\phi_{\M}^q(x)$, the LFWF is derived straightforwardly;  the produced physical picture would be valid at $\zeta_H$. Given the robustness of the SDE formalism to compute PDAs, we shall employ predictions obtained within this framework as model inputs~\cite{Cui:2020tdf,Ding:2015rkn}. The specific set of PDAs we consider is the following ($\bar{x}=1-x$):
\begin{eqnarray}
\nonumber
\phi_\pi^u(x)&=&20.226\, x\bar{x}\,[1-2.509 \sqrt{x\bar{x}}+2.025x\bar{x}]\;,\\
\phi_K^u(x)&=&18.04\,x\bar{x}\,[1+5x^{0.032}\bar{x}^{0.024}-5.97x^{0.064}\bar{x}^{0.048}]\;, \nonumber\\
\phi_{\eta_c}^c(x)&=&9.222\, x\bar{x}\,\text{exp}\,[-2.89(1-4x\bar{x})]\;, \nonumber\\
\phi_{\eta_b}^b(x)&=&12.264\,x\bar{x}\,\text{exp}[-6.25(1-4x\bar{x})]\;.\label{eq:PDAsSDE}
\end{eqnarray}
The expressions above properly capture our contemporary knowledge of such distributions, namely, the soft endpoint behavior and the dilation/compression with respect to the asymptotic distribution~\cite{Lepage:1979zb}:
\begin{equation}
    \label{eq:PDAasym}
    \phi_{asy}(x) = 6x(1-x)\;.
\end{equation}
As can be seen in Fig.~\ref{fig:PDA1}, pion and kaon PDAs are dilated with respect to $\phi_{asy}(x)$, while those containing heavy quarks are narrower. As noted for the kaon, the asymmetry between the $s$ and $u$-quark masses produces a skewed distribution, while the rest of the PDAs are symmetrical. 

The remaining ingredients are the parameter $\nu$ and the constituent masses $M_q$. Regarding the former, $\nu=1$ is a natural choice since it yields the correct asymptotic behavior of the BSWF~\cite{Roberts:1994dr}. 
Concerning the values of the constituent masses, we shall employ available experimental~\cite{Zyla:2020zbs},  SDE~\cite{Chang:2013nia,Eichmann:2019bqf,Bhagwat:2006xi,Miramontes:2021exi,Raya:2022ued} and lattice QCD~\cite{Dudek:2007zz,Dudek:2006ej} results on the charge radii as benchmarks, and determine $M_{q}$ via Eq.~\eqref{eq:crgeneral}. Table~\ref{tab:params} collects the constituent quark masses that define our AM and the corresponding charge radii.

\begin{table}[ht]
\centering
\caption{Model inputs: meson and quark masses (in GeV). $M_q$ values are fixed via Eq.~\eqref{eq:crgeneral} using the quoted charge radii. In the case of $\eta_c$ and $\eta_b$, we quote $r_{\M}^q=r_{\M}^{\bar{h}}$ rather than $r_{\M}$, which is strictly zero. The list of distribution amplitudes entering the relevant equations are found in Eq.~\eqref{eq:PDAsSDE}. }
\label{tab:params}
\begin{tabular}[t]{c|c|l||c|c}
\hline
Meson\;\; & $m_{\M}$ \;\;& \;$r_{\M}$ (in fm) \;& Quark\;\; & $M_q$ \;\;\\
\hline
$\pi^+$ & 0.14 &\; 0.659~\cite{Zyla:2020zbs,Chang:2013nia}\; & $u$ & 0.317\\
$K^+$ & 0.49 &\; 0.600~\cite{Eichmann:2019bqf,Miramontes:2021exi,Raya:2022ued} \;& $s$ & 0.574\\
$\eta_c$ & 2.98 &\; 0.255~\cite{Dudek:2007zz,Dudek:2006ej}\; & $c$ & 1.65\\
$\eta_b$ & 9.39  &\; 0.088~\cite{Bhagwat:2006xi}\; & $b$ & 5.09\\
\hline
\end{tabular}
\end{table}%

With the AM fully determined, the produced LFWFs are shown in Fig.~\ref{PionLFWF}. It is clear that the heavier mesons exhibit a much slower damping as $k_\perp^2$ increases. Furthermore, just as the PDAs, the LFWFs as a function of $x$ are found to be more compressed in this case.

The valence-quark GPDs are then obtained appealing to the overlap representation of the LFWF, Eqs.~(\ref{GPDdefinition},\ref{eq:GPDfinal}). Pion and kaon results are shown in the bottom panel of Fig.~\ref{fig:Pion-Kaon-GPDs}, while those of $\eta_c$ and $\eta_b$ can be found in Fig.~\ref{EtacGPD}. The GPDs for the heavier mesons naturally have a narrower profile along the $x$-axis and are harder along the $-t$-axis. Moreover, the upper panel of Fig.~\ref{fig:Pion-Kaon-GPDs} also displays a comparison between the GPDs obtained directly from Eq.~\eqref{eq:GPDfinal} and the approximate representation of~\eqref{eq:GPDLFHQCD}. The derived valence quark PDFs are found in Fig.~\ref{PDFs}. As one would expect from Eq.~\eqref{eq:defPDF}, the characteristic features exhibited by the PDAs, of dilation and narrowness, are filtered into PDFs. To emphasize it, we notice that the plots in the above mentioned figure display the scale-free parton-like profile
\begin{equation}
    \label{eq:PDFsf}
    q_{sf}(x)=30 x^2(1-x)^2\;.
\end{equation}
Given our preferred value $M_s \approx 1.8 \, M_u$, the $s$-in-$K$ momentum fraction at the hadronic scale is $\textless x;\zeta_H \textgreater_K^s = 0.55$, about $4\%$ larger than typical values~\cite{Cui:2020tdf,Cui:2020dlm}. The pion and kaon PDFs are then evolved from the hadronic scale, $\zeta_H=0.33(1\pm 0.1)$ GeV, to the experimentally accessible scale of $\zeta_5:=5.2$ GeV. The evolution procedure is detailed, for instance, in Refs.~\cite{Raya:2021zrz,Rodriguez-Quintero:2019fyc}. Fig.~\ref{PDFsE} displays the outcome. In the top panel of this figure, the valence quark as well as gluon and sea quark pion PDFs are shown. At the evolved scale, we find typical values of momentum fraction distribution in pion~\cite{Cui:2020tdf,Cui:2020dlm}: $\textless x;\zeta_H \textgreater_\pi^{\text{val}} = 0.41(4)$, $\textless x;\zeta_H \textgreater_\pi^{\text{sea}} = 0.14(2)$, $\textless x;\zeta_H \textgreater_\pi^{\text{glue}} = 0.45(3)$. The bottom panel of Fig.~\ref{PDFsE} compares the valence quark PDFs in pion and kaon. Then again, our choice of $M_s$ produces a slightly larger momentum fraction for the $s$ valence-quark at such scale, $\textless x; \zeta_5 \textgreater_K^s = 0.25$, and a smaller one for the $u$ quark, $\textless x; \zeta_5 \textgreater_K^u = 0.17$. Concerning the large-$x$ exponents of the valence quark distributions, we find that
\begin{equation}
    u_{\pi,K}(x\to 1;\zeta_5) \sim (1-x)^{ \beta_{eff}}\;,\;\beta_{eff} \approx 2.8,
\end{equation}
where $\beta_{eff}$ must be interpreted as an effective exponent rather than that obtained from the known evolution equations of $\beta(\zeta_H)$~\cite{Cui:2020tdf, RuizArriola:2004ui}. Moreover, the $x$-domain of applicability and interpretation of $\beta(\zeta_H)$ is not without its ambiguities and requires special care~\cite{Courtoy:2020fex}.
The electromagnetic FFs are displayed in Figs.~(\ref{Pion-Kaon-FF}, \ref{EtacFF}). As can be noted therein, pion and kaon FFs agree with the available experimental data~\cite{AMENDOLIA1986168,JeffersonLabFpi:2000nlc,JeffersonLabFpi-2:2006ysh} and previous SDE calculations~\cite{Chang:2013nia,Eichmann:2019tjk}. The $\eta_c$ FF is compared with lattice QCD~\cite{Dudek:2006ej,Dudek:2007zz} and SDE results in the contact interaction (CI) model~\cite{Bedolla:2016yxq}. Similarly, the $\eta_b$ result is contrasted with CI model results and with previous determinations with an AM for heavy quarkonia~\cite{Raya:2017ggu}. Both $\eta_c$ and $\eta_b$ form factors show a satisfactory compatibility with earlier reliable predictions.

The IPS-GPDs are derived from the  approximate LFHQCD-inspired parametrization of the GPD, introduced in~\cite{deTeramond:2018ecg} and quoted in Eq.~\eqref{eq:GPDLFHQCD}. For illustrative purposes, we have considered the convenient  representation of Eq.~\eqref{eq:bqblack}, which produces the pion and kaon results shown in Fig.~\ref{PionIP}. The quark region is identified with $x\textgreater 0$, while the antiquark lies in the $x\textless 0$ domain.  The symmetry in the pion case is a natural consequence of the isospin symmetry, whereas the contraction on the $s$-in-$K$ distribution is a result of $M_s$ being larger than $M_u$. In fact, as the constituent quark mass becomes larger, it is expected that the quark plays an increasingly major role in determining the center of transverse momentum; furthermore, the distributions become narrower and the maximums become larger. Given the compact representation of the IPS-GPDs, the values $(x^\text{max},\,b_\perp^\text{max})_{\M}^q$ where $\textbf{b}_{\M}^q(x,b_\perp)$ acquires its global maximum, can be readily identified: 
\begin{equation}
    b_\perp^\text{max}=\sqrt{2 \hat{f}_{\M}^q(x^\text{max})}\;,
\end{equation}
and $x^\text{max}$ is the real-valued solution of
\begin{equation}
    q_{\M}(x) f'(x) -2 q_{\M}'(x) f(x) = 0\;.
\end{equation}
It is thus clear that a constant PDF yields the point particle limit $(|x^\text{max}|,\,b_\perp^\text{max})_{\M}^q \to (1,0)$. The location of the maximum and its value are reported in Table~\ref{tab:ImpParMax} for different mesons. Finally, according to Eq.~\eqref{eq:aveMSTE} and our model inputs, we report the expectation values of the MSTE for the kaon:
\begin{equation}
    \textless b_\perp^2 \textgreater_{\M}^u = 0.76\, r_K^2\;,\;\textless b_\perp^2 \textgreater_{\M}^s = 0.47\, r_K^2\;;
\end{equation}
while for the heavy quarkonia and pion in isospin symmetric case we can infer the result from Eq.~\eqref{eq:expMSTE}. 

\begin{table}[ht]
\centering
\caption{Global maximum $\mathcal{I}_{\M}^q:=\text{max}[\textbf{b}_{M}^q(x,\,b_\perp)]$ and its location $(x^{\text{max}},\,b_\perp^{\text{max}}/r_{\M})$. }
\label{tab:ImpParMax}
\begin{tabular}[t]{c||c|c||c|c}
\hline
Meson & $\;(x^{\text{max}},\,b_\perp^{\text{max}}/r_{\M})_{\M}^q\;$ & $\;\;\mathcal{I}_{\M}^q\;\;$ & $\;(x^{\text{max}},\,b_\perp^{\text{max}}/r_{\M})_{\M}^{\bar{h}}\;$ & $\;\;\mathcal{I}_{\M}^{\bar{h}}\;\;$ \\
\hline
$\pi$ & (0.90, 0.10) & 3.19 & (-0.90, 0.10) & 3.19 \\
$K$   & (0.76, 0.18) & 2.03 & (-0.88, 0.14) & 4.79 \\
$\eta_c$ & (0.53, 0.56) & 3.99 & (-0.53, 0.56) & 3.99 \\
$\eta_b$ & (0.52, 0.60) & 4.90 & (-0.52, 0.60) & 4.90 \\
\hline
\end{tabular}
\end{table}%
\noindent For the pion and kaon, one can visually verify
these tabulated results in Fig.~(\ref{PionIP}). 
This completes the presentation of computed results. 

\section{Summary and scope}

In this article, we put forward a fairly general AM for the pseudoscalar meson BSWF, which preserves 
its primary attractive feature of guaranteeing most calculations continue to be analytic. For systematic and visual clarity, we italicize its main features
and our key results as follows:

\begin{itemize}

\item The key functions of the model are the spectral density $\rho_{\rm M}(w)$ and  $\Lambda(w)$, which play the defining role for the dominant BSA,~Eq.~(\ref{Anzats2}), of the pseudo-scalar mesons we study.

\item The function $\Lambda(w)$, defined through Eq.~(\ref{Lambda}), is quadratic in $w$ which is as high as we can go in the power of this polynomial while still preserving the analytic nature of the calculations involved. In all previous models,
$\Lambda(w)$ was merely taken as a constant mass scale $\Lambda$.

\item

Allowing $\Lambda$ to become a function of the variable $w$ allows us to connect LFWF with PDA algebraically, Eq.~(\ref{eq:LFWFPDArel}), without having the need to rather arbitrarily concoct the spectral density. 

\item Despite having emphasized the previous point, the fact remains that the spectral density can be extracted unequivocally through the knowledge of the PDA. 

\item Given the most up to date pseudo-scalar meson PDAs, we merely need to fix $\nu$ and $M_{q,\bar{h}}$. As $\nu=1$ is a natural choice, we can safely say that the quark mass is the only free parameter to fix the model.

\item 
Crucially, the measure of factorizability of $x$ and $k_{\perp}$ in the LFWF is evident through Eqs.~\eqref{Lambda} and~\eqref{eq:LFWFPDArel}. An immediate consequence is the hadronic scale relation between the PDA and the PDF, Eq.~\eqref{eq:defPDF}. This factorization is completely reinstated in the chiral limit, thus reproducing known results~\cite{Chouika:2017rzs, Mezrag:2016hnp, Mezrag:2014jka,Chavez:2021koz,Chavez:2021llq} as a particular case. 

\item With the exception of the leading-twist PDAs (which are external inputs) and the charge radii (used as benchmarks to set the values of the constituent masses, $M_q$), the rest of the distributions and other quantities derived herein are predictions. 
\end{itemize}

Notably, our ingenuous model faithfully reproduces previously known results concerning pions~\cite{Raya:2022eqa,Raya:2021zrz,Zhang:2021mtn}. Our findings for the kaon are slightly different from those reported  therein but can readily and correctly be attributed to the larger strange quark mass favored by this model. However, the description of pion and kaon is  compatible with our experimental understanding of these mesons. It is worth mentioning that our pion valence-quark PDF is also compatible with the results from Ref.~\cite{dePaula:2022pcb}, in which the authors also obtain a NIR for the BSWF, but through the resolution of the corresponding BSE (modeling some of the ingredients that go into the latter). Novel results employing sophisticated mathematical techniques also validate the NIR approach~\cite{Eichmann:2021vnj}. The distributions reported for $\eta_c$ and $\eta_b$, and other related quantities, are a completely novel feature of our study. In general, when a comparison is possible, our results also show agreement with other theoretical treatments such as SDEs, lattice QCD, as well as with experimental results. The structure of $\pi^0,\,\eta_c$ and $\eta_b$ is currently being investigated within this model, via two photon transition form factors. In the future we expect to carry out kindred studies of mesons with heavy-light quarks as well as adapt the procedure to  study the entangled $\eta-\eta'$ system and eventually baryons. 

\section{Acknowledgements}

LA acknowledges the FAPESP
postdoctoral fellowship grant no. 2018/17643-5.
The research of LA and IMH. was also supported by CONACyT through the national fellowship program. AB wishes to thank CIC of UMSNH for the research grant 4.10.


\section*{Appendix: $\rho (y)$ Differential equation}

\begin{figure}[!htb]
    \centering
    \includegraphics[scale=.24]{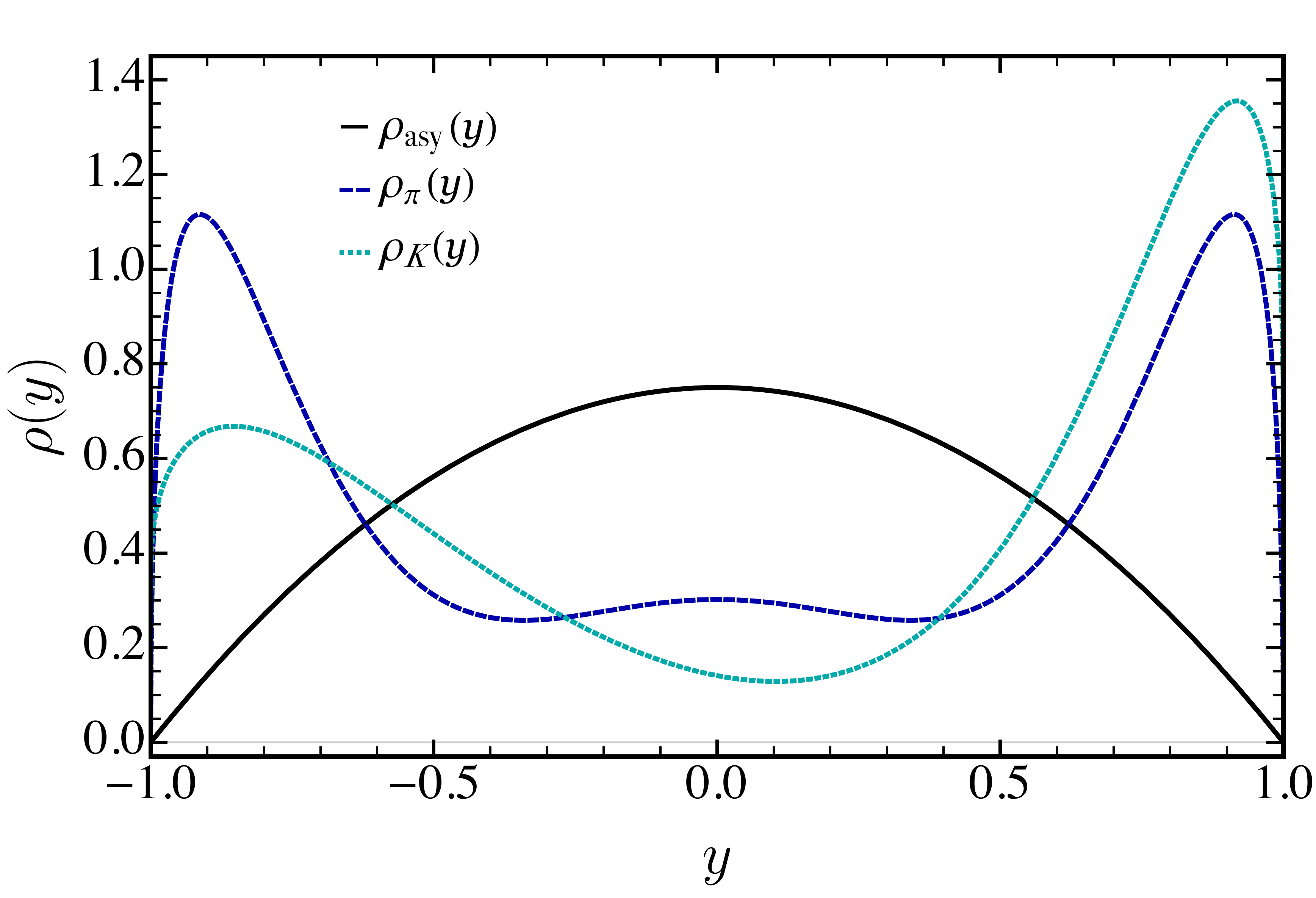}
    \caption{
    Spectral density for the pion (dashed-blue line), the kaon (dashed-cyan line) and the one produced by $\phi_{asy}(x)$ in the chiral limit. We fix $\nu=1$ for the three cases. The parameters chosen correspond to the ones in Table \ref{tab:params}.}
    \label{fig.:SpectralDensity}
\end{figure}

From Eqs. (\ref{eq:FM1},\ref{LFWF1}-\ref{eq:PDAdefinition}), it is possible to derive the relation between the PDA and the spectral density $\rho_M$:
\begin{eqnarray}
\varphi(y)  &=& \frac{ 1 }{ 2\nu F_N }  \left[ \int_{-1}^{y} \hspace{-2mm} dw \left( \hspace{-1mm} \frac{ 1-y }{ 1-w } \hspace{-1mm} \right)^{\nu} \hspace{-1.5mm} + \hspace{-1mm} \int_{y}^{1} \hspace{-2mm} dw \left( \hspace{-1mm} \frac{ 1+y }{ 1+w } \hspace{-1mm} \right)^{\nu}  \right] \Tilde{\rho}_{\M}^{\, \nu}(w) \nonumber \\
&& \hspace{1.0cm} \times  \frac{\left[(1+y)M_q + (1-y)M_{\bar{h}} \right] }{\Lambda^{2\nu}_y} \,,
\end{eqnarray}
where the variable $y=1-2x$ has been introduced and we have used the definitions $\varphi(y) \equiv \phi_M^q(\frac{1}{2}(1-y))$ and $F_N=\frac{4}{3} \pi^2 f_{\M} n_{\M}$. The above integral equation can be inverted to a differential equation by differentiating three times, with respect to $y$, and summing up the resulting equations. This procedure yields an expression for $\rho_{\M}$ in terms of derivatives of $\varphi$:
\begin{eqnarray}
\eta_N \, \rho_{\M}(y) = \lambda_{\nu}^{(2)}(y) \varphi''(y)+\lambda_{\nu}^{(1)}(y) \varphi'(y)+\lambda_{\nu}^{(0)}(y) \varphi(y) \,, \nonumber \\
\end{eqnarray}
where $\eta_N$ is a normalization factor such that 
\begin{eqnarray*}
\int_{-1}^1 \rho_{\M}(y) dy = 1
\end{eqnarray*} 
while the other quantities are
\begin{eqnarray}
\lambda_{\nu}^{(2)}(y) &=& -\frac{1-y^2}{\chi_{+}} \,, \\
\lambda_{\nu}^{(1)}(y) &=& 2 \frac{\nu y}{\chi_{+}} - 2 \frac{\chi_{-}}{\chi_{+}^2} + \frac{\nu \chi_{-}}{\Lambda_{y}^2}\,, \\
\lambda_{\nu}^{(0)}(y) &=& \big\{ 2 \nu \chi_{+}^2 \Lambda_y^2 \left( \chi_{+}^2 - 2 \left( 1 + (1 - \nu) y^2 + \nu \right) \Lambda_y^2 \right)  \nonumber \\
&& \, + 4 y (1 - \nu) \left( 2 \Lambda_y^2 - \nu \chi_{+}^2 \right) \Lambda_y^2 \chi_{+} \chi_{-} \nonumber \\
&& \, - \left( \nu (1 - \nu) \chi_{+}^4 + 2 \nu \chi_{+}^2 \Lambda_y^2 - 8 \Lambda_y^4 \right) \chi_{-}^2 \big\} / \Theta_y \,, \nonumber \\
\end{eqnarray}
with the definitions $\chi_{\pm} = (1-y) M_{\bar{h}} \pm (1+y) M_{q}$ and $\Theta_y = -4 \left( 1 - y^2 \right) \chi_{+}^3 \Lambda_y^4$. By setting $\nu = 1$, $\lambda_{\nu}^{(1,0)}$  are reduced to
\begin{eqnarray}
\lambda_{1}^{(1)}(y) &=& 2 \frac{y}{\chi_{+}} - 2 \frac{\chi_{-}}{\chi_{+}^2} + \frac{\chi_{-}}{\Lambda_{y}^2}\,, \\
\lambda_{1}^{(0)}(y) &=& - \frac{ \left( \chi_{+}^2 - 4 \Lambda_y^2 \right) \left( \chi_{+}^2 - \chi_{-}^2 \right) }{2 \left( 1 - y^2 \right) \chi_{+}^3 \Lambda_y^2} \,.
\end{eqnarray}
Furthermore, in the chiral limit:
\begin{eqnarray}
\lambda_1^{(2)}=-\frac{(1-y^2)}{2M_q}\;,\;\lambda_1^{(1)}=\lambda_1^{(0)}=0\;,
\end{eqnarray}
ensuring that our model recovers known result~\cite{Chang:2013nia}: 
\begin{eqnarray}
 \phi_{\M}^q(x)=\phi_{asy}(x)&=&6x(1-x)\\ &\Longleftrightarrow&\; \rho_{\M}(w)=\rho_{asy}(w):=\frac{3}{4}(1-w^2)   \;.\nonumber
\end{eqnarray}
Beyond the chiral limit, but still keeping the most natural choice $\nu=1$, the corresponding  pion and kaon spectral densities are plotted in Fig.~\ref{fig.:SpectralDensity}. The input PDAs, parametrized according to Eqs.~\eqref{eq:PDAsSDE}, are displayed in the upper panel of Fig.~\ref{fig:PDA1}. 

Though for the purposes of this work (namely computing LFWFs, GPDs and distributions derived therefrom) the determination of $\rho_{\M}$ is not required at all, it is worth stressing that the AM we have introduced enables a straightforward derivation of the spectral density from the prior knowledge of the PDA, thus avoiding the need of assuming a particular {\em ad hoc} representation for $\rho_{\M}$. This shall be useful for future explorations that require the explicit knowledge of the BSWF, and hence of the spectral density.

\newpage


\bibliography{main}

\end{document}